# NORMAL MODES OF THE *d* > 1 FERMION GAS

Daniel C. Mattis*

Dept. of Physics and Astronomy,

University of Utah

Salt Lake City, UT 84112 USA

Abstract:

In the present work we solve the many-fermion problem in $d > 1$ dimensions by identifying its normal modes. Even when fermions are subject to internal interactions they continue to share a Fermi sea. We start by decomposing the Fermi sea into sectors labeled by wave-vectors $\vec{q}$, each containing a set of orthogonal, independent normal modes. Operators that create or destroy normal modes within any given sector are shown to commute with those originating in other sectors even in the presence of two-body forces. Each sector's Hamiltonian is a quadratic form in its normal modes and can be diagonalized in a multitude of ways. One convenient method, elaborated here and obviating the need for many-body perturbation theory, consists of mapping it onto a $d = 1$ *generalized* Luttinger model – a solvable model that is a functional equivalent of a *boson string*. The product of exactly calculable eigenstates, one from each sector, is an exact eigenstate of the full many-body problem. We examine bosonization of fermions, *exchange* and *pairing* forces and quasiparticles, all from this point of view.

*mattis@physics.utah.edu

**(0) INTRODUCTION.** The principal purpose of the present work is easily stated: it is to show that a gas of fermions sustains *normal modes*, just as does any extended molecule or solid, whether initially comprised of classical or even of *quantum* particles. After quantizing these normal modes we show how to diagonalize their Hamiltonians. The work is directly based on publications that recently generalized the Luttinger model of interacting fermions,[1,2] although it significantly expands and improves on them. We shall show that in any dimension, a momentum-conserving two-body interaction that perturbs the ground state of $N$ fermions can be separated into non-interacting sectors labeled by the internal momenta $\vec{q}$. We examine the sectors to identify the ingredients of a satisfactory solution of the many-body problem.

Already starting in the 1970's and 1980's there have been numerous recorded attempts to generalize Luttinger's exactly solvable one-dimensional model to dimensions $d > 1$, including an early study in 2D by the present author[3] in connection with then-emergent high-$T_c$ superconductivity. Typically, such generalizations required the Fermi surface to possess a discrete number of flattened portions. A rigorous mathematical justification for this approach was recently spelled out by De Woul and Langmann.[4] But the present theory differs from its predecessors[1,2,3,4] insofar as it decomposes an arbitrarily shaped Fermi sea into an *infinite* number of infinitesimal sectors. The basis operators within each sector commute with those in all other sectors. Each sector maps onto a *generalized* Luttinger model characterized by a string of bosons.[1] There ensues a set of coefficients $A_n^m$ tailored to a given dispersion function $e(\vec{k})$; the two-body interactions are grafted on afterwards, in a canonical fashion.

Below, we provide explicit derivations and calculations while also addressing ancillary concerns. These center on the nature of *exchange forces* and on the role played by particle-conservation within each sector. We also address such issues as bosonization of spin waves, virtual Cooper pairs and the connection to Landau's Fermi liquid.

The exposition has been rendered as comprehensive and self-contained as possible. Numbered sections bear descriptive titles and all nontrivial results are either proved or illustrated by examples, while equations or derivations are repeated – even if they are redundant – wherever needed for maximum clarity.

**(1) ORIGINAL FORMULATION.** Consider the prototype conservative system of $N$ ($N \to \infty$) SU(2) fermions in a volume $V = L^d$ ($L^d \to \infty$) in dimension $d > 1$ at finite particle density $N/V$. Its particles interact *via* two-body forces, of which the Coulomb repulsion is one example, at temperatures low compared to the degeneracy temperature. The use of periodic boundary conditions mandates momentum conservation at each collision. The total Hamiltonian describing this conservative system is the sum of kinetic energy $H_1$ and interactions $H_2$. Using anticommuting fermion operators $c_\sigma^\dagger(\vec{k})$ and $c_\sigma(\vec{k})$ that create or destroy fermions labeled by $(\vec{k}, \sigma)$, subject to anti-commutation relations, *e.g.*, $c_\sigma^\dagger(\vec{k})c_{\sigma'}(\vec{k}') + c_{\sigma'}(\vec{k}')c_\sigma^\dagger(\vec{k}) \equiv \{c_\sigma^\dagger(\vec{k}), c_{\sigma'}(\vec{k}')\} = \delta_{\sigma,\sigma'}\delta_{\vec{k},\vec{k}'}$, we express kinetic energy $H_1$ as:

$$H_1 = \sum_k \sum_\sigma \varepsilon(\vec{k})c_\sigma^\dagger(\vec{k})c_\sigma(\vec{k}) = \sum_k \sum_\sigma \varepsilon(\vec{k})\tilde{n}_\sigma(\vec{k}) \qquad (1)$$

where $\varepsilon(\vec{k}) = e(\vec{k}) - e_F$ is the energy measured from the Fermi level ($e_F$); $\sigma$ (actually, $\sigma_z$) refers to spin "up" or "down". To fix ideas we assume an "effective mass" approximation[5] $e(\vec{k}) = \hbar^2 k^2 / 2m*$ and units $\dfrac{\hbar^2}{2m*} = 1$ and $k_F = 1$. The Fermi level is chosen such that ½$N$ states have $\varepsilon(\vec{k}) < 0$.

In the ground state of $N$ noninteracting (free) fermions each occupation-number operator $\tilde{n}_\sigma(\vec{k}) = c_\sigma^\dagger(\vec{k})c_\sigma(\vec{k})$ has eigenvalue 1 if $\varepsilon(\vec{k}) < 0$ and 0 for $\varepsilon(\vec{k}) > 0$. This discontinuity in eigenvalue of $\tilde{n}_\sigma(\vec{k})$, from 1 just below the Fermi level to 0 just above it diminishes, on average, in the presence of two-body interactions. It also disappears at any finite temperature $T$. Thermal fluctuations and two-body interactions are qualitatively similar, inasmuch as both cause low-lying elementary excitations to become enmeshed into the dynamical fabric, depleting states below $e_F$ of particles while causing formerly empty states above $e_F$ to become occupied. [6]

Two-body interactions are governed by an expression that is quartic in fermions,

$$
\begin{aligned}
H_2 &= \frac{1}{2!Vol} \sum_{\vec{k},\vec{k}',\vec{q}} V(\vec{q}) \sum_{\sigma,\sigma'} c_\sigma^\dagger(\vec{k}+\frac{\vec{q}}{2}) c_{\sigma'}^\dagger(\vec{k}'-\frac{\vec{q}}{2}) c_{\sigma'}(\vec{k}'+\frac{\vec{q}}{2}) c_\sigma(\vec{k}-\frac{\vec{q}}{2}) \\
&= \frac{1}{2!Vol} \sum_{\vec{q}} V(\vec{q}) : \left\{ \sum_{\sigma'} \sum_\sigma \rho_{\sigma'}(\vec{q}) \rho_\sigma(-\vec{q}) \right\} :
\end{aligned}
\tag{2}
$$

where $\rho_\sigma(\vec{q}) = \sum_{\vec{k}} c_\sigma^\dagger(\vec{k}+\frac{\vec{q}}{2}) c_\sigma(\vec{k}-\frac{\vec{q}}{2})$ and colons : ... : indicates that the product of fermion operators in the curly brackets is to be "normally ordered" (annihilation operators to the right, creation operators to the left.) In both the illustrations and in the examples below, we specialize to just the following: $V(\vec{q}) = U = $ constant for $q < q_0$, zero otherwise. This point-like *Hubbard interaction* is governed by two independent parameters: the coupling constant $U$ and the cutoff $q_0$, a maximal wave-vector inversely proportional to the lattice spacing $d_0$.

Although the interaction $H_2$ in Eq. (2) is relevant to all the properties of the Fermi gas it is invoked only late in the paper. The first and most important task is to introduce a new algebra to reëxpress the *kinetic energy*, Eq. (1), along with the *Fermi sea and its low-lying excitations*, in terms of a complete set of boson normal modes in 2D and 3D. In this we follow a template originally crafted for the analysis of Luttinger's model in 1D.[1]

**(2) THE FERMI SEA AND ITS BOSONIC EXCITATIONS.** Classically, collective vibrations of individual particles about presumed positions of equilibrium are known as "normal modes." After being orthogonalized and quantized, the associated creation or annihilation operators satisfy Bose-Einstein commutation relations *ab initio*. Notable examples include photons in electrodynamics, the phonons of solid state physics and magnons in Heisenberg's model of magnetism.

But here the premises are different. For, starting from what is already highly quantum-mechanical *fermion* physics, we propose to construct linear combinations of low-lying excitations that *behave* as bosons – but then only in the "thermodynamic limit." (This jargon means, essentially, that both *N*, the number of degrees of freedom and *L*, the characteristic length of the system, both tend to $\infty$ in such a way that $N/L^d$ remains finite in *d* dimensions.) The system's dynamics are governed by the Hamiltonian that is shown to be quadratic in these bosons. Once *H* is diagonalized, its eigenstates yield a complete picture of ground state and excited states.

Everyone knows of the *plasma mode* of the charged electron gas or the *zero sound* of an uncharged fluid. It is less well known that the Fermi gas is capable of supporting an infinite number of additional, distinct, collective excitations. After identifying them as quantized density- and spin-fluctuations, or virtual propagating Cooper-pairs (those related to superconductivity,) we explore possible paths to exact diagonalization.

To start (omitting spin indices at first, for typographical simplicity) let us now define the bare-bones prototype *annihilation* operator $a_0(\vec{q})$ as the followings linear combination of elementary excitations of the free (noninteracting) Fermi gas:

$$a_0(\vec{q}) \equiv \frac{C(\vec{q})}{\sqrt{q \times Vol}} \sum_{\vec{k}} \theta(\vec{k} \cdot \vec{q}) c^\dagger(\vec{k} - \vec{q}/2) c(\vec{k} + \vec{q}/2) \ . \tag{3}$$

Here and throughout, $\theta(\vec{k} \cdot \vec{q})$ is the Heaviside function ($\theta = 1$ if its argument is positive, zero otherwise.) $Vol = L^d$ in *d* dimensions. Although no single constituent of Eq. (3), $c^\dagger(\vec{k} - \vec{q}/2) c(\vec{k} + \vec{q}/2)$, satisfies Bose-Einstein commutation relations, *all* do lower the momentum and kinetic energy of any state to which they are applied so $a_0(\vec{q})$ has to be a "lowering" operator of sorts. We then postulate that the *commutator* $[a_0(\vec{q}), a_0^\dagger(\vec{q})] \equiv$ $a_0(\vec{q})a_0^\dagger(\vec{q}) - a_0^\dagger(\vec{q})a_0(\vec{q})$ is equal to 1. Actually this "normalization" requirement is what *determines* the parameter $C(\vec{q})$ in (3). Thus,

$$1 = [a_0(\vec{q}), a_0^\dagger(\vec{q})] = \frac{C^2(\vec{q})}{|\vec{q}|Vol} \sum_k \theta(\vec{k} \cdot \vec{q})(\tilde{n}(\vec{k} - \vec{q}/2) - \tilde{n}(\vec{k} + \vec{q}/2))$$

$$= \frac{C^2(\vec{q})}{|\vec{q}|(2\pi)^d} \int d^d k \, \theta(\vec{k} \cdot \vec{q})(\tilde{n}(\vec{k} - \vec{q}/2) - \tilde{n}(\vec{k} + \vec{q}/2))$$

$$= \frac{C^2(\vec{q})}{|\vec{q}|(2\pi)^d} \int d^d k \, \theta(\vec{k} \cdot \vec{q})(f(\vec{k} - \vec{q}/2) - f(\vec{k} + \vec{q}/2)) \qquad (4)$$

In the last line of (4) we replaced the *rhs* of the equation (initially the sum of operators) by its thermodynamic average. Although it is not obvious that we may do this, *it is not an approximation.*[7]

The distribution function *f* is the average of $\tilde{n}(\vec{k}) = c^\dagger(\vec{k})c(\vec{k})$. It is a function of its argument, of the temperature *T* and of the interactions. For this substitution to be valid, the average $f(\vec{k}) = <\tilde{n}(\vec{k})>$ must be taken not just with respect to thermal fluctuations but must also be subject to *all* internal dynamical forces (interactions). We return toward the end of this work to the actual calculation of $f(\varepsilon)$. The calculations and the figure suggest that at low temperatures ($kT < e_F/10$) the normalization parameter $C^2(\vec{q}) \simeq 20$ is insensitive to $q$ and to increased rounding of *f* at the Fermi surface. These results are also indicative of what we should expect at, or near, *T*=0, in the event the internal forces $H_2 \neq 0$ became finite (but insufficiently strong to erase the Fermi surface.) Calculations in 3D show similar tendencies.

Actually, $a_0(\vec{q})$ is just the first in an infinite set ("field") of similarly constituted bosons, $a_j(\vec{q}) \equiv \frac{C(\vec{q})}{\sqrt{q \times Vol}} \sum_k \theta(\vec{k} \cdot \vec{q})\Phi_j(\vec{k}, \vec{q})c^\dagger(\vec{k} - \vec{q}/2)c(\vec{k} + \vec{q}/2)$, whose "wavefunctions" $\Phi_j$ (typically real) are chosen from among a complete orthogonal set at each fixed $q$ and obey $1 = \dfrac{C^2(\vec{q})}{|q|(2\pi)^d} \int d^d k \, \theta(\vec{k} \cdot \vec{q}) \Phi_j^2(\vec{k}, \vec{q})(f(\vec{k} - \vec{q}/2) - f(\vec{k} + \vec{q}/2))$,

allowing the previously introduced *normalization* requirement, $[a_j(\vec{q}), a_j^\dagger(\vec{q})] = 1$, to be satisfied.

As will be shown later, the functions $\Phi_j(\vec{k}, \vec{q})$ have to have a certain number of "radial" nodes[8] (that is, more or less parallel to $\vec{q}$) or "transverse" nodes (more or less perpendicular to $\vec{q}$) as function of $\vec{k}$ to ensure the $a_j$'s satisfy the Bose-Einstein algebra:

$$[a_i(\vec{q}), a_j^\dagger(\vec{q}')] = \delta_{i,j} \delta_{\vec{q},\vec{q}'} \text{ and of course, } [a_i(\vec{q}), a_j(\vec{q}')] = [a_j^\dagger(\vec{q}'), a_i^\dagger(\vec{q})] = 0 . \quad (5)$$

*Within* each sector, the $a_0$ operator is the only one in each set of $a_j$'s that is *nodeless* ($\Phi_0 = 1$ is obviously so!) That is but one reason it is singled out. Now let us look at the commutation relations (5) in somewhat more detail.

*Two Operators in Two Distinct Sectors*. Eqs. (5) include the requirement that *any two* annihilation operators commute. (Also, that two creation operators commute.) Let us prove this first by direct calculation of two $a_0$ operators culled from *two distinct* sectors; then the generalization to arbitrary $a_j$'s from two distinct sectors should become obvious.

$$[a_0(\vec{q}), a_0(\vec{q}')] \equiv \dfrac{C(\vec{q})C(\vec{q}')}{\sqrt{q \times q' \times Vol^2}} \sum_k \sum_{k'} \theta(\vec{k} \cdot \vec{q}) \theta(\vec{k}' \cdot \vec{q}') \times$$

$$\left[ c^\dagger(\vec{k} - \vec{q}/2) c(\vec{k} + \vec{q}/2), \ c^\dagger(\vec{k}' - \vec{q}'/2) c(\vec{k}' + \vec{q}'/2) \right]$$

$$= \dfrac{C(\vec{q})C(\vec{q}')}{\sqrt{q \times q' \times Vol^2}} \sum_k \sum_{k'} \theta(\vec{k} \cdot \vec{q}) \theta(\vec{k}' \cdot \vec{q}') \times$$

$$\{ \delta_{\vec{k}+\vec{q}/2,\vec{k}'-\vec{q}'/2} c^\dagger(\vec{k} - \vec{q}/2) c(\vec{k}' + \vec{q}'/2) - \delta_{\vec{k}'+\vec{q}'/2,\vec{k}-\vec{q}/2} c^\dagger(\vec{k}' - \vec{q}'/2) c(\vec{k} + \vec{q}/2) \}$$

This looks impossibly messy. But after regrouping the terms, redefining dummy indices of summation and performing some trivial algebra, one gets:

$$[a_0(\vec{q}), a_0(\vec{q}')] =$$

$$\frac{1}{\sqrt{Vol}} \frac{C(\vec{q})C(\vec{q}')}{\sqrt{qq'}Vol} \sum_{\vec{k}} \left( \theta((k - \frac{\vec{q}'}{2}) \cdot \vec{q}) \theta((k + \frac{\vec{q}}{2}) \cdot \vec{q}') - \theta((k - \frac{\vec{q}}{2}) \cdot \vec{q}') \theta((k + \frac{\vec{q}'}{2}) \cdot \vec{q}) \right)$$

$$\times c^\dagger(k - \frac{q+q'}{2}) c(k + \frac{q+q'}{2})$$

$$= \frac{1}{\sqrt{Vol}} \left( A_i a_i(\vec{q} + \vec{q}') + A_j a_j^\dagger(-\vec{q} - \vec{q}') \right) \qquad (6)$$

in which the operators $a_i$ and $a_j^\dagger$ on the *rhs* of (6) are boson operators defined in the related sectors $\pm(\vec{q} + \vec{q}')$; they assumed normalized but are otherwise unspecified. The coefficients $A_i$ and $A_j$ are both finite (real but not necessarily positive) functions and are expressed in the units of length. In the special case $\vec{q} + \vec{q}' = 0$ these coefficients are both identically zero,[9] hence $[a_0(\vec{q}), a_0(-\vec{q})] \equiv 0$ *identically*. In another trivial special case, $\vec{q} = \vec{q}'$, the summand in (6) *also* vanishes identically. Other than these special cases, the *rhs* of (6) also vanishes – but only *in the thermodynamic limit*. That is, both quantities $A_i$ and $A_j$ are finite, as are $<a_i^\dagger a_i>$ and $<a_j^\dagger a_j>$, while $1/\sqrt{Vol} \to 0$. Therefore the product (6) vanishes in the limit.

These results generalize, *mutatis mutandis*, to the full set of $a_j$ operators culled from any two distinct sectors. *We* assume, therefore, that $[a_i(\vec{q}), a_j(\vec{q}')] = 0$ for all $i, j$.

However, we distinguish the *exact* commutation relations[8] $[a_i(\vec{q}), a_j(-\vec{q})] \equiv 0$ from the *weaker* commutation relations $[a_i(\vec{q}), a_j(-\vec{q}')] \propto \dfrac{\text{operator}}{L^{d/2}} \to 0$ satisfied only in the thermodynamic limit (and only under the supposition that the "operator" in the numerator remain finite in this limit.) Such a distinction might be viewed as just so much hair-splitting in the context of the many-body problems – whether metal physics or nuclear matter. But in *small* systems, *e.g.*, atoms, nuclei, small molecules or quantum dots, this distinction suggests caution before applying the theory without appropriate modification!

*Commutators of 2 Operators Within any Given Sector*. Generalizing Eq. (6), one shows that two annihilation operators inhabiting a common sector must also commute, in the sense that, $[a_i(\vec{q}), a_j(\vec{q})] \propto \dfrac{(\text{finite operator})_{i,j}}{L^{d/2}} \to 0$ .

Certainly if $i=j$ , the "operator" in the numerator vanishes *identically*. A second rule, that applies at $i=j$ and earlier was denoted the "normalization condition," *i.e.*,

$$[a_j(\vec{q}), a_j^\dagger(\vec{q})] = 1 = \frac{C^2(\vec{q})}{|q|(2\pi)^d} \int d^d k\, \theta(\vec{k} \cdot \vec{q}) \Phi_j^2(\vec{k}, \vec{q})(f(\vec{k} - \vec{q}/2) - f(\vec{k} + \vec{q}/2))$$  , serves to

determine the magnitude of $C(q)$ – if not the functional form of $\Phi_j$ .

More generally, when $i \neq j$ within a fixed sector, *all* commutators must vanish, not just $[a_i(\vec{q}), a_j(\vec{q})]$ . Even such quantities as,

$$[a_i(\vec{q}), a_j^\dagger(\vec{q})] = \frac{C^2(\vec{q})}{|q|(2\pi)^d} \int d^d k\, \theta(\vec{k} \cdot \vec{q}) \Phi_i(\vec{k}, \vec{q})\, \Phi_j(\vec{k}, \vec{q})(f(\vec{k} - \vec{q}/2) - f(\vec{k} + \vec{q}/2)) \quad (7)$$

*must vanish*. We recognize Eq. (7) as the requirement for the "orthogonality" of distinct operators. *A special case*: the integral (7) also vanishes if either $a_i$ or $a_j$ is replaced by $a_0$ .

Therefore, given any $j \neq 0$ ,

$$[a_0(\vec{q}), a_j^\dagger(\vec{q})] = \frac{C^2(\vec{q})}{|q|(2\pi)^d} \int d^d k\, \theta(\vec{k} \cdot \vec{q}) \Phi_j(\vec{k}, \vec{q})(f(\vec{k} - \vec{q}/2) - f(\vec{k} + \vec{q}/2)) = 0 \quad (8A)$$

and $[a_0(\vec{q}), a_j(\vec{q})] = \dfrac{\text{finite operator}}{L^{d/2}} \to 0$ . \hfill (8B)

The second equation, (8B), tells us nothing we did not already know. But in order that (8A) vanish it is necessary that the corresponding $\Phi_j$ have 1 or more nodes (*i.e.*, changes of sign) as a function of $\vec{k}$ , given that the other factor $\theta(\vec{k} \cdot \vec{q})(f(\vec{k} - \vec{q}/2) - f(\vec{k} + \vec{q}/2))$ in the integrand is positive semi-definite. This nodal structural requirement was stated earlier without proof. Also, we must distinguish *transverse* and *radial* nodes.

Let us denote $\Phi_j$'s without a transverse node *s*-waves, those with one transverse node *p*-waves, 2 transverse nodes *d*-waves, etc. Depending on the particulars of $H_2$ there may exist a "good" quantum number $l$ that counts the number of transverse nodes ($l$=0 for *s*-waves, etc.,) and a quantum number $n$ counting radial nodes just as in atomic physics.

The vanishing of Eq. (7) suggests that two distinct $\Phi_j$'s within a common sector will not have the same number of both radial and transverse nodes. Here – as in atomic physics – whenever two orthogonal wave functions occupy a common domain and have the same number of transverse nodes they must have differing numbers of radial nodes, or *vice-versa*.

**(3) DECOMPOSITION OF $H$ INTO NONINTERACTING SECTORS.** Although it may seem obvious, the following statement is both seminal and crucial in the further elaboration of the theory:

Whenever the Hamiltonian can be decomposed into a sum over independent sectors, labeled by $q$, $l$, or whatever, such that each sector is diagonalizable independently of the others, the many-body problem is thereby "reduced to quadrature."[10] Then, the global wave function is merely the product of eigenstates plucked from each sector.

Let the following serve as an example of such factorization. We let $H_1$ be a surrogate for the kinetic energy and $H_2$ for a momentum-conserving two-body interaction, expressing both in some hypothetically complete set of bosons $b_j$ similar to the $a_j$ introduced earlier, all labeled by $q$ in one (arbitrarily chosen) hemisphere.

Let $H_1 = \sum_{\vec{q}, q_z > 0} H_1(\vec{q})$, where $H_1(\vec{q}) = \sum_{i,j} \{L_{i,j}(\vec{q}) b_i^\dagger(\vec{q}) b_j(\vec{q}) + L_{i,j}(-\vec{q}) b_i^\dagger(-\vec{q}) b_j(-\vec{q})\}$ and

similarly, $H_2 = \sum_{\vec{q}, q_z > 0} H_2(\vec{q})$ with $H_2(\vec{q}) = \sum_{i,j} \{M_{i,j}(\vec{q}) b_i(\vec{q}) b_j(-\vec{q}) + H.c.\}$. Then,

$$H = \sum_{\vec{q}, q_z > 0} (H_1(\vec{q}) + H_2(\vec{q})) = \sum_{\vec{q}, q_z > 0} H(\vec{q}), \qquad (9)$$

That is, in each sector,

$$H(\vec{q}) = L_{i,j}(\vec{q}) b_i^\dagger(\vec{q}) b_j(\vec{q}) + L_{i,j}(-\vec{q}) b_i^\dagger(-\vec{q}) b_j(-\vec{q}) + (M_{i,j}(\vec{q}) b_i(\vec{q}) b_j(-\vec{q}) + H.c.)$$

$= $ using $\{1 + 2 + 3 + 4\}$ as a short-hand for the 4 terms.

We wish to prove that $H(\vec{q})$ and $H(\vec{q}')$ commute, hence that $H$ is *separable*; implying that all its eigenstates can be expressed as product states over all distinct sectors $\vec{q}$.

Given that $H(\vec{q})$ and $H(\vec{q}')$ are each quadratic in bosons, the proof that they commute requires the vanishing of $[H(\vec{q}), H(\vec{q}')] = [\{1 + 2 + 3 + 4\}, \{1', 2', 3', 4'\}]$, *i.e.*,

$$\Big[ \Big( \{L_{i,j}(\vec{q}) b_i^\dagger(\vec{q}) b_j(\vec{q}) + L_{i,j}(-\vec{q}) b_i^\dagger(-\vec{q}) b_j(-\vec{q})\} + \{M_{i,j}(\vec{q}) b_i(\vec{q}) b_j(-\vec{q}) + H.c.\} \Big),$$

$$\Big( \{L_{i,j}(\vec{q}') b_i^\dagger(\vec{q}') b_j(\vec{q}') + L_{i,j}(-\vec{q}') b_i^\dagger(-\vec{q}') b_j(-\vec{q}')\} + \{M_{i,j}(\vec{q}') b_i(\vec{q}') b_j(-\vec{q}') + H.c.\} \Big) \Big] = 0.$$

$$(10)$$

We can prove the stronger result that, not only does (10) vanish, but so does *every one* of the 16 individual commutators that comprise it.

Take one of the 16 commutators, say $[1, 3'] = [AB, CD] = 0$, where $A = b_i^\dagger(\vec{q})$, $B = b_j(\vec{q})$, $C = b_i(\vec{q}')$ and $D = b_i(-\vec{q}')$ (omitting the irrelevant $L$ or $M$ coefficients). It is evaluated with the help of an iterative commutator identity,

$$[AB, CD] \equiv A[B, CD] + [A, CD]B \equiv A\{[B, C]D + C[B, D]\} + \{[A, C]D + C[A, D]\}B.$$

$A$ and $B$ each are labeled $\bar{q}$, and $C$, $D$ are labeled $\pm \bar{q}\,'$. We established in Eqs. (6) and (7) that whenever two such sectors are truly distinct *all* commutators involving the synthetic bosons (whether $[b_i(\bar{q}\,'),b_j(\bar{q})]=0$ or $[b_i(\bar{q}\,'),b_j^{\dagger}(\bar{q})]=0$) take the form $\dfrac{b \text{ or } b^{\dagger}}{L^{d/2}}$ and must vanish in the thermodynamic limit. So, $[AB,CD]$ decomposes into the sum of $A[B,C]D, AC[B,D],[A,C]DB$ and $C[A,D]B$, *each of which* is the product of three momentum-conserving operators, all presumably of O(1), situated in three distinct sectors ($q,q'$, and $\pm q \pm q'$). Division by $L^{d/2}$ causes each of these quantities to vanish in the limit.

This conclusion holds for each of the 16 terms in the expansion of Eq.(10), *QED*.

Next we shall *derive* just this sort of quadratic form $H(\bar{q})$ for what were originally interacting fermions. It shall be viewed as the Hamiltonian of a one-dimensional harmonic string. Harmonic strings can be diagonalized by a variety of techniques.[11] Once the string's eigenstates are obtained explicitly, the complete set of eigenstates of the *total* $H$ are formed from products of eigenstates, one from each of the individual sectors. Also, all *extensive* properties –total energy, entropy, etc. – are sums of the respective quantities calculated within individual sectors.

But there is a fly in the ointment: if *any* matrix element of $H$ connected separate sectors (such as "backward scattering" does in the one-dimensional Luttinger model,) its very existence would destroy the separability of $H$ and vitiate the product state solutions. In an Appendix we assess plausible inter-sector matrix elements to assess whether their importance rises to the level of their nuisance factor.

**(4) DENSITY, SPIN AND COOPER-PAIRING CHANNELS.** To simplify the exposition we omitted the spin degrees of freedom in the preceding. But the very definition of sectors depends on the interaction Hamiltonian $H_2$. Originally quartic in the fermion operators, its decomposition into bilinear forms of bosons depends very much on the spins. Therefore let us now specialize to fermions with spin, SU(2) particles, examples of which are the electrons or holes that live in metals or the neutrons and protons of nuclear physics. The simplest example of such a system is the Hubbard model.[12] We use it in this paper to illustrate first, the separation into sectors and second, the methods of solution.

Elsewhere, in other publications, we turn to more realistic interactions such as the Coulomb repulsion, which are physically more relevant but call for far more sophisticated and complicated analysis.

Using $\varrho_o$ as defined just after Eq. (2), the Hubbard model interaction Hamiltonian is written,

$$H_2 = \frac{U}{Vol} \sum_{\bar{q}, q_z > 0} \left\{ \rho_\uparrow(\bar{q})\rho_\downarrow(-\bar{q}) + \rho_\downarrow(\bar{q})\rho_\uparrow(-\bar{q}) \right\} \qquad . \qquad (11)$$

Terms such as $\rho_\uparrow(\bar{q})\rho_\uparrow(-\bar{q})$ and $\rho_\downarrow(\bar{q})\rho_\downarrow(-\bar{q})$ are conspicuously absent because $V(q)=U=$ constant in this model, hence all "direct" interactions among particles of the same spin are precisely canceled by "exchange" terms. They need not be included – even *ab initio*.[13]

The summand in (11) can be decomposed into bilinear forms of density-density interactions and Heisenberg-like interactions, each couched in its respective bosons. The procedure starts by rewriting terms in curly brackets in the above equation,

$$\rho_\uparrow(\bar{q})\rho_\downarrow(-\bar{q}) + \rho_\downarrow(\bar{q})\rho_\uparrow(-\bar{q}) =$$
$$\frac{1}{2}(\rho_\uparrow(\bar{q}) + \rho_\downarrow(\bar{q}))(\rho_\uparrow(-\bar{q}) + \rho_\downarrow(-\bar{q})) \; - \; \frac{1}{2}(\rho_\uparrow(\bar{q}) - \rho_\downarrow(\bar{q}))(\rho_\uparrow(-\bar{q}) - \rho_\downarrow(-\bar{q})) \qquad (12)$$

The first parentheses (with (+)'s) yields the density-density interactions that we denote $H_{2,dir}$. We'll start with that. The second parentheses (with (–)) yields just *one* of three components of the spin-spin exchange interactions – as we shall see below.

This reformulation in $a_0$ comes about after comparing Eqs. (2) and (3) to identify density operators $\varrho$ with the $a_0$ operators: $a_{0,\sigma}(\vec{q}) + a_{0,\sigma}^\dagger(-\vec{q}) = \frac{C(\vec{q})}{\sqrt{q \times Vol}} \rho_\sigma(\vec{q})$ assuming $C(\vec{q})$ is rotationally invariant to spin orientation $\sigma$. The total particle density is then,

$$\rho_\downarrow(\vec{q}) + \rho_\uparrow(\vec{q}) = \sqrt{\frac{q \times Vol}{C^2(\vec{q})}} \Big( a_{0,\downarrow}(\vec{q}) + a_{0,\uparrow}(\vec{q}) + a_{0,\downarrow}^\dagger(-\vec{q}) + a_{0,\uparrow}^\dagger(-\vec{q}) \Big)$$
$$\equiv \sqrt{\frac{2 \times q \times Vol}{C^2(\vec{q})}} \Big( a_0(\vec{q}) + a_0^\dagger(-\vec{q}) \Big) \qquad (13)$$

In arriving at the second line we define a linear combination $a_{0,\downarrow}(\vec{q}) + a_{0,\uparrow}(\vec{q}) \equiv \sqrt{2}\, a_0(\vec{q})$. The quadratic form $\frac{1}{2}(\rho_\uparrow(\vec{q}) + \rho_\downarrow(\vec{q}))(\rho_\uparrow(-\vec{q}) + \rho_\downarrow(-\vec{q}))$ yields the direct density-density (*alias* charge-charge) interaction, and upon being reëxpressed in the normalized $a_0$ boson operators we have become familiar with, is

$$H_{2,dir} = U \sum_{\vec{q},\, q_z > 0} \frac{|q|}{C^2(q)} \Big\{ a_0^\dagger(\vec{q}) a_0(\vec{q}) + a_0^\dagger(-\vec{q}) a_0(-\vec{q}) + \Big( a_0(\vec{q}) a_0(-\vec{q}) + H.c. \Big) \Big\}. \qquad (14)$$

Repeat the procedure with $\rho_\downarrow(\vec{q}) - \rho_\uparrow(\vec{q}) = \sqrt{\frac{2 \times q \times Vol}{C^2(\vec{q})}} \Big( z_0(\vec{q}) + z_0^\dagger(-\vec{q}) \Big)$, these being a new set of operators, the "longitudinal" *spin-densities* $z_0$. If we assume that $f(\vec{k})$ and $C(\vec{q})$ are independent of spin orientation $\sigma$, $z_0$ too is normalized. It is,

$$z_0(\vec{q}) \equiv \frac{C(\vec{q})}{\sqrt{2 \times q \times Vol}} \sum_k \theta(\vec{k} \cdot \vec{q}) \Big( c_\uparrow^\dagger(\vec{k} - \vec{q}/2) c_\uparrow(\vec{k} + \vec{q}/2) - c_\downarrow^\dagger(\vec{k} - \vec{q}/2) c_\downarrow(\vec{k} + \vec{q}/2) \Big) \quad (15)$$

and $[z_0(\vec{q}), z_0^\dagger(\vec{q})] = 1$ by analogy with the $a_0$. Its contribution to $H_2$ is,

$$H_{2,z} = -U \sum_{\vec{q},\, q_z > 0} \frac{|q|}{C^2(q)} \Big\{ z_0^\dagger(\vec{q}) z_0(\vec{q}) + z_0^\dagger(-\vec{q}) z_0(-\vec{q}) + \Big( z_0(\vec{q}) z_0(-\vec{q}) + H.c. \Big) \Big\}. \qquad (16)$$

But this can only be part of an "exchange" Hamiltonian. Symmetry requires two additional bosons, related to $z_0$ by $90^0$ spatial rotations of the Cartesian coordinate axes:

$$x_0(\vec{q}) \equiv \frac{C(\vec{q})}{\sqrt{2 \times q \times Vol}} \sum_k \theta(\vec{k} \cdot \vec{q}) \left( c_\uparrow^\dagger(\vec{k} - \vec{q}/2) c_\downarrow(\vec{k} + \vec{q}/2) + c_\downarrow^\dagger(\vec{k} - \vec{q}/2) c_\uparrow(\vec{k} + \vec{q}/2) \right) \quad (17)$$

and

$$y_0(\vec{q}) \equiv \frac{C(\vec{q})}{i\sqrt{2 \times q \times Vol}} \sum_k \theta(\vec{k} \cdot \vec{q}) \left( c_\uparrow^\dagger(\vec{k} - \vec{q}/2) c_\downarrow(\vec{k} + \vec{q}/2) - c_\downarrow^\dagger(\vec{k} - \vec{q}/2) c_\uparrow(\vec{k} + \vec{q}/2) \right) \quad (18)$$

Their algebra is of interest. The operators $a_0$, $x_0$, $y_0$ and $z_0$, are constructed as *energy-* and *momentum*-lowering operators hence these operators are not Hermitean. This set of operators commute among themselves but not with the set of their Hermitean conjugate operators: *e.g.,* $[z_0(\vec{q}), z_0^\dagger(\vec{q})] = 1$ as we have seen. More significantly, mixed commutators

$$[z_0(\vec{q}), x_0^\dagger(\vec{q})] = iy_0(\vec{0}) \frac{C(\vec{q})}{\sqrt{2 \times q \times Vol}} \qquad , \qquad (19)$$

and various permutations, also might fail to vanish.

For in the presence of symmetry-breaking or long-range ferromagnetic order, one or more of the quantities $x_0(0), y_0(0)$, or $z_0(0)$ might assume a macroscopic value $\geq \sqrt{Vol}$ instead of O(1). In that case, the *rhs* of (19) is no longer negligible and the commutation relations among the $x_0$, $y_0$ and $z_0$ operators and their Hermitean conjugates would satisfy a different sort of Lie algebra. Also, the coefficients $C(\vec{q})$ could depend on spin orientation, etc. The same is true in the event of *anti*ferromagnetic order, in which case one needs to search nonvanishing commutators such as, $[z_0(\vec{q}), x_0^\dagger(\vec{q} + \vec{Q})]$, for a $\vec{Q}$ that is representative of some antiferromagnetic ordering. The question is, whether any of $x_0(\vec{Q}), y_0(\vec{Q})$, or $z_0(\vec{Q})$ can assume a macroscopic value $\geq \sqrt{Vol}$ (which would imply some sort of nonzero long-range order.)

Relegating such complications to future investigations let us assume for present purposes that *there is no symmetry breaking of any kind and no macroscopic magnetization*. Then $x_0(\vec{q}), y_0(\vec{q})$ and $z_0(\vec{q})$ and $x_0^\dagger(\vec{q}) x_0(\vec{q})$ etc., like $a_0(\vec{q})$ and $a_0^\dagger(\vec{q}) a_0(\vec{q})$, all remain operators of magnitude O(1) at all $q$. Under those circumstances the three spin components together with the density fluctuation operators $a_0$ constitute an orthonormal set of bosons.[14]

When assembling all the interactive normal modes embedded in $H_2$, we find not just $H_{2,dir}$ and $H_{2,z}$ as in the above construction, but additionally $H_{2,x}$ and $H_{2,y}$. Although these new contributions are required by symmetry they could also have been obtained directly by pairing fermions of spins "up" with spins "down" in the original quartic expression of $H_2$. (Such "transverse" terms, although required by symmetry, were mistakenly omitted in early formulations of the Luttinger model.[1])

Here is a detailed derivation. Indicating pairing by underlining, in Eq. (11) one finds not just terms $+\underline{c_\uparrow^\dagger(\vec{k}+\vec{q}/2) c_\uparrow(\vec{k}-\vec{q}/2)}\,\underline{c_\downarrow^\dagger(\vec{k}'-\vec{q}/2) c_\downarrow(\vec{k}'+\vec{q}/2)}$ leading to Eq. (14) but *additionally*: $-\underline{c_\uparrow^\dagger(\vec{k}+\vec{q}/2) c_\downarrow(\vec{k}-\vec{q}/2)}\,\underline{c_\downarrow^\dagger(\vec{k}'-\vec{q}/2) c_\uparrow(\vec{k}'+\vec{q}/2)}$. This last is obtained by exchanging the two annihilation operators and redefining the dummy indices of summation. (These manipulations are only possible because $V(q) = U$ is constant in this model.) Upon combining *all* such exchange terms that are required by rotational symmetry we obtain a rotationally invariant spin-dependent interaction. *Specifically* for the Hubbard model, it is:

$$H_{2,\sigma} = -U \sum_{\vec{q}, q_z > 0} \frac{|q|}{C^2(q)} \left\{ \vec{\sigma}_0^\dagger(\vec{q}) \cdot \vec{\sigma}_0(\vec{q}) + \vec{\sigma}_0^\dagger(-\vec{q}) \cdot \vec{\sigma}_0(-\vec{q}) + \left( \vec{\sigma}_0(\vec{q}) \cdot \vec{\sigma}_0(-\vec{q}) + H.c. \right) \right\} \quad (20)$$

The components of the vector spin operator $\vec{\sigma}_0(\vec{q}) = (x_0(\vec{q}), y_0(\vec{q}), z_0(\vec{q}))$ that appears here are those that were spelled out in Eqs. (15), (17) and (18). This vector-spin exchange Hamiltonian, Eq. (20), replaces the longitudinal exchange Hamiltonian in Eq. (16).

Note that the negative coupling constant $-U$ favors the creation of spontaneously nonvanishing magnitudes of $\vec{\sigma}_0(\vec{q})$ not just at some discrete $q = 0$ or $Q$, but in *all* sectors. If this appears to challenge the original premise of *no* spontaneous magnetization or long-range order, it is only because we have not yet considered $H_1$.

The energy of freely moving particles *always* has its minimum energy when $<\vec{\sigma}_0(\vec{q})> \equiv 0$ for all $\vec{q}$. Although a delicate balancing of two competing tendencies may be the result, let us assume provisionally that the kinetic energy prevails and there is no long-range order.

But before proceeding along this track we remark on *yet another* pairing allowed by Eq. (11), of type $+ \underbrace{c_\uparrow^\dagger(\vec{k} + \vec{q}/2) c_\downarrow^\dagger(-\vec{k} + \vec{q}/2)}\underbrace{c_\downarrow(\vec{k}' + \vec{q}/2) c_\uparrow(-\vec{k}' + \vec{q}/2)}$. First introduced in the BCS theory,[15] this is generally known as "Cooper pairing." Its inclusion with the 3 spin- pairings and the density-pairing adds up to the 5 degrees of freedom at the heart of the SO(5) mean-field theories of high-$T_c$ superconductivity (popular at one time.[16]) Here, however, the positive coupling constant $+U$ stifles formation of the energy-gap producing Cooper pairing (also known as "off-diagonal" long range order) associated with superconductivity. Nevertheless, regardless of the sign of $U$, the ground state energy benefits from the zero-point contribution of virtual Cooper pairings or "*cooperons*."

Looking further into the cooperons one finds their designation to be not quite as straightforward as one would like them to be, for they apparently exist as *two* distinct species – each compatible with the above pairing. The lowering operators of the first, labeled (+), consist of the following linear combinations of elementary excitations:

$$\chi_{j,+}(\vec{q}) = \frac{D_+(\vec{q})}{\sqrt{Vol}} \sum_{\vec{k}} \theta\Big(\varepsilon(\vec{k}+\vec{q}/2)+\varepsilon(-\vec{k}+\vec{q}/2)\Big)\Psi_{j,+}(\vec{q}\,|\,\vec{k})c_\downarrow(\vec{k}+\vec{q}/2)c_\uparrow(-\vec{k}+\vec{q}/2)$$

(21A)

that act only on states *outside* the joint Fermi sea of two particles ($\vec{k}$'s for which $\varepsilon(\vec{k}+\vec{q}/2)+\varepsilon(-\vec{k}+\vec{q}/2) > 0$, hence the subscripts "+"). They *lower* the total momentum by $q$ and also *lower* the total energy by an amount which, while variable, is always positive. The second set of operators act only *inside* the joint Fermi sea (note $c_\sigma^\dagger$). They too lower the total momentum by $q$ while lowering the total energy by a variable but always positive amount.

The subscript "–" indicates $\varepsilon(\vec{k}+\vec{q}/2)+\varepsilon(-\vec{k}+\vec{q}/2) < 0$. The relevant operators are:

$$\chi_{j,-}(\vec{q}) = \frac{D_-(\vec{q})}{\sqrt{Vol}} \sum_{\vec{k}} \theta\Big(-\varepsilon(\vec{k}-\vec{q}/2)-\varepsilon(-\vec{k}-\vec{q}/2)\Big)\Psi_{j,-}(\vec{q}\,|\,\vec{k})c_\uparrow^\dagger(-\vec{k}-\vec{q}/2)c_\downarrow^\dagger(\vec{k}-\vec{q}/2)$$

(21B)

Together the $\chi_\pm$'s form a complete, complementary set for the cooperons. Their respective Hermitean conjugates are energy- *and* momentum-*raising* operators,

$$\chi_{j,+}^\dagger(\vec{q}) = \frac{D_+(\vec{q})}{\sqrt{Vol}} \sum_{\vec{k}} \theta\Big(\varepsilon(\vec{k}+\vec{q}/2)+\varepsilon(-\vec{k}+\vec{q}/2)\Big)\Psi_{j,+}^*(\vec{q}\,|\,\vec{k})c_\uparrow^\dagger(-\vec{k}+\vec{q}/2)c_\downarrow^\dagger(\vec{k}+\vec{q}/2)$$

(22A)

and

$$\chi_{j,-}^\dagger(\vec{q}) = \frac{D_-(\vec{q})}{\sqrt{Vol}} \sum_{\vec{k}} \theta\Big(-\varepsilon(\vec{k}-\vec{q}/2)-\varepsilon(-\vec{k}-\vec{q}/2)\Big)\Psi_{j,-}^*(\vec{q}\,|\,\vec{k})c_\downarrow(\vec{k}-\vec{q}/2)c_\uparrow(-\vec{k}-\vec{q}/2)$$

(22B)

Although the choice $\pm\Big(\varepsilon(\vec{k}+\vec{q}/2)+\varepsilon(-\vec{k}+\vec{q}/2)\Big)$ of argument of the Heaviside functions may seem arbitrary, it serves to define the distinct sectors uniquely and leads verifiably to the correct normalization: $[\chi_\alpha, \chi_\beta^\dagger] = +\delta_{\alpha,\beta}$.[17] Just like the $\Phi$'s, the $\Psi$'s are, or can be chosen, real in most applications. Their magnitude is determined, as it was for the $\Phi_j$'s before, by the normalization. Their dependence on $q$ and $k$ depends, as we shall see, on the functional form of the kinetic energy. Note that neither of the initial wave functions $\Psi_{0,\pm} = 1$ depends on $k$ at all (except for the necessary cutoff at large $k$ !)

Commutators of any two $\chi$ operators from distinct sectors vanish. Moreover *in the thermodynamic limit* the $\chi$'s commute with *all* other bosons – whether these deal with spin or with charge fluctuations, whether or not their $q$'s are the same. But because $D_+$ depends explicitly on the *u-v* cutoff $q_0$ while $D_-$ depends on $k_F$, these two constants are necessarily unequal.[18] By symmetry, $D_\pm(-\vec{q}) = D_\pm(+\vec{q})$.

We express the contribution to $H_2$ of the two species of cooperons in Hubbard's model as a bilinear form in the $\chi$'s, using symmetry (that is,),

$$H_{2,\chi} = +U\sum_{\vec{q}}\left(\frac{\chi_{o,-}(\vec{q})}{D_-(\vec{q})} + \frac{\chi_{o,+}^\dagger(-\vec{q})}{D_+(\vec{q})}\right) \cdot \left(\frac{\chi_{o,-}^\dagger(\vec{q})}{D_-(\vec{q})} + \frac{\chi_{o,+}(-\vec{q})}{D_+(\vec{q})}\right) \qquad . \qquad (23)$$

Here, we recall, $\Psi_{0,\pm} = 1$ is the wave function in each of these operators, evidently nodeless.

**(5) CONSTRUCTION OF A STRING THEORY.** To calculate the restoring forces that prevent $\vec{\sigma}_0$ from growing arbitrarily large, and to complete the decomposition into non-overlapping normal modes, one needs first to transform the kinetic energy – initially a bilinear form in fermions, $H_1 = \sum_k \sum_\sigma \varepsilon(\vec{k}) c_\sigma^\dagger(\vec{k}) c_\sigma(\vec{k})$ – into a bilinear form in *bosons*. These are then mapped onto strings. This procedure of "bosonization" is achieved solely using $H_1$.[19] Mathematically, it consists of a mapping of fermionic states onto the Hilbert space of the low-lying, elementary, excitations.

We start by deriving the equations of motion that create the subsequent operators, *e.g.*, $\{a_1(\vec{q}), a_2(\vec{q}), \ldots\}$. The equations of spin densities are algebraically identical to those of the density operators hence the calculation need only be performed once. Equations of motion of the cooperons are also similar, although their coefficients differ due to the distinct geometry of their sectors. In any event, each set of normal modes maps onto its own dedicated string.

We shall observe that the "density" kinetic energy $H_1$ is isomorphic to a quadratic form in the set of operators $\{a_0(\vec{q}), a_1(\vec{q}), a_2(\vec{q}), \ldots\}$ and their Hermitean conjugates. This is proved *by construction*. The coefficients of this quadratic form depend explicitly on the functional form of the fermions' dispersion $e(\vec{k})$ and on the dimension $d$, and are determined by an operator version of the *Lanczös procedure* – an algorithm that is well-known in linear algebra and a staple of every modern computer software library, wherein square matrices are reduced to tridiagonal form to facilitate the task of mapping them onto 1D or calculating their eigenvalues.

The set starts with the nodeless $a_0(\vec{q})$ first seen in Eq. (3). Together with $a_0^\dagger(\vec{q})$ it is the only operator that appears in $H_{2,dir}$. We use it to construct the first in a sequence of commutator "equations of motion," *viz.*,

$$[a_0(\vec{q}), H_1] = \frac{C(\vec{q})}{\sqrt{q \times Vol}} \sum_k \theta(\vec{k} \cdot \vec{q}) \left( \varepsilon(\vec{k} + \vec{q}/2) - \varepsilon(\vec{k} - \vec{q}/2) \right) c^\dagger(\vec{k} - \vec{q}/2) c(\vec{k} + \vec{q}/2) \quad (24)$$

If we were studying Luttinger's original 1D model by this method, the parenthesis $\left( \varepsilon(\vec{k} + \vec{q}/2) - \varepsilon(\vec{k} - \vec{q}/2) \right) \propto \pm q$ would be seen[20] to be independent of $k$; hence the *rhs* of (24) would be seen to be proportional to $a_0$. One could immediately conclude there is closure, as the string reduces to a point. But, more generally, in 2D or 3D, or even in 1D, there is "dispersion." Then the *rhs* needs to be decomposed into *two* parts: the first proportional to the initial $a_0(\vec{q})$ while the second is proportional to some new operator we shall denote $a_1(\vec{q})$. Of this second operator nothing is known except it is presumed normalized, and also "orthogonal" to $a_0(\vec{q})$ – by which is meant that $a_1(\vec{q})$ and its Hermitean conjugate $a_1^\dagger(\vec{q})$ both commute with $a_0(\vec{q})$. That said, it is possible to recast (24) and its conjugate equation in a generic, more perspicuous, form:

$$[a_0(\vec{q}), H_1] = A_0^0(\vec{q})a_0(\vec{q}) + A_0^1(\vec{q})a_1(\vec{q}) \ , \tag{25A}$$

$$[H_1, a_0^\dagger(-\vec{q})] = A_0^0(\vec{q})a_0^\dagger(-\vec{q}) + A_0^1(\vec{q})a_1^\dagger(-\vec{q}) \ . \tag{25B}$$

The $A$'s are coefficients, the subscripts of which refer to the rank of the equation and the superscripts to the position within that equation. They can be assumed real and symmetric, $i.e.$ $A_0^1(\vec{q}) = A_1^0(\vec{q})$, mainly because $H_1$ itself is both real and Hermitean.

The following describes the procedure by which the coefficients are calculated. First, $A_0^0(\vec{q})$ is isolated with the aid of the commutator of (25) with $a_0^\dagger(\vec{q})$,

$$[[a_0(\vec{q}), H_1], a_0^\dagger(\vec{q})] = [A_0^0(\vec{q})a_0(\vec{q}) + A_0^1(\vec{q})a_1(\vec{q}), \ a_0^\dagger(\vec{q})] = A_0^0(\vec{q})$$
$$= \frac{C^2(\vec{q})}{|q|Vol}\sum_k \theta(\vec{k}\cdot\vec{q})\Big(\varepsilon(\vec{k} + \vec{q}/2) - \varepsilon(\vec{k} - \vec{q}/2)\Big)\Big(f(\vec{k} - \vec{q}/2) - f(\vec{k} + \vec{q}/2)\Big) \tag{26}$$

The first line of (26) makes explicit use of orthogonality $[a_1(\vec{q}), a_0^\dagger(\vec{q})] = 0$ and of normalization $[a_0(\vec{q}), a_0^\dagger(\vec{q})] = 1$, while the second line shows the explicit calculation using the fermions' anticommutator relations and subjected to the Central Limit Theorem (which calls for replacing the number operators $c^\dagger(\vec{k})c(\vec{k}) = \tilde{n}(\varepsilon(\vec{k}))$) by their thermodynamic averages $<\tilde{n}(\varepsilon(\vec{k})>_{TA} = f(\varepsilon(\vec{k}))$ when being summed.)[7]

Once we know $A_0^0(\vec{q})$ it is a simple matter to obtain $A_0^1(\vec{q})$ and then set up recursion equations for all successive $A$'s. But before proceeding to this next step let us show how this procedure can be conceptually and algebraically simplified, using a more elegant and symbolic notation that is exhibited next.

**(6) AVERAGES.** Let us make use of a *nonnegative*, *normalized* distribution function,

$P_{\vec{q}}(\vec{k}) = \dfrac{C^2(\vec{q})}{|q|Vol}\theta(\vec{k}\cdot\vec{q})\Big(f(\vec{k}-\vec{q}/2)-f(\vec{k}+\vec{q}/2)\Big)$, defined in each density sector $q$. The

average of an arbitrary function $G$ subjected to this probability is,

$$<G(\vec{k})> \equiv \frac{C^2(\vec{q})}{|q|Vol}\sum_k \theta(\vec{k}\cdot\vec{q})\Big(G(\vec{k})\Big)\Big(f(\vec{k}-\vec{q}/2)-f(\vec{k}+\vec{q}/2)\Big). \tag{27}$$

The average is generally a function of $q$ but, to keep the notation simple, we don't exhibit this.

Suppose we set $G = 1$, then Eq. (27) yields an identity 1=<1> for all $q$, admittedly just a neat restatement of what we had earlier dubbed the "normalization condition."

With $\Delta\varepsilon \equiv \varepsilon(\vec{k}+\vec{q}/2)-\varepsilon(\vec{k}-\vec{q}/2)$, Eq. (26) for $A_0^0(\vec{q})$ can now be written more succinctly as,

$$A_0^0(\vec{q})=<\varepsilon(\vec{k}+\vec{q}/2)-\varepsilon(\vec{k}-\vec{q}/2)> \equiv <\Delta\varepsilon>. \tag{28}$$

This quantity is plotted in Fig. 1 as function of $q$ and seen to vanish as $|q|$ in the long-wavelength limit. For the calculation of $A_0^1(\vec{q})$, let us rearrange (25) as follows:

$A_0^1(\vec{q})a_1(\vec{q})=[a_0(\vec{q}),H_1]-A_0^0(\vec{q})a_0(\vec{q})$, *i.e.*,

$$A_0^1(\vec{q})a_1(\vec{q})=$$
$$\frac{C(\vec{q})}{\sqrt{|q|Vol}}\sum_k \theta(\vec{k}\cdot\vec{q})\Big(\varepsilon(\vec{k}+\vec{q}/2)-\varepsilon(\vec{k}-\vec{q}/2)-A_0^0(\vec{q})\Big)c^\dagger(\vec{k}-\vec{q}/2)c(\vec{k}+\vec{q}/2) \tag{29}$$
$$=\frac{C(\vec{q})}{\sqrt{|q|Vol}}\sum_k \theta(\vec{k}\cdot\vec{q})\Big(\Delta\varepsilon-<\Delta\varepsilon>\Big)c^\dagger(\vec{k}-\vec{q}/2)c(\vec{k}+\vec{q}/2)$$

Without knowing anything more about $a_1$ except that it is normalized, we can evaluate the commutator of (29) with its own Hermitean conjugate to obtain $\left|A_0^1(\vec{q})\right|^2$. That is, $[A_0^1(\vec{q})a_1(\vec{q}), A_0^1(\vec{q})^* a_1^\dagger(\vec{q})] = \left| A_0^1(\vec{q}) \right|^2$. Extraction of the positive root, $A_0^1(\vec{q})$, allows it to be identified as the *variance* in the distribution of the elementary excitations $\Delta\varepsilon$,

$$A_0^1(\vec{q}) = \sqrt{<\Delta\varepsilon^2> - <\Delta\varepsilon>^2} = \sqrt{<(\Delta\varepsilon - <\Delta\varepsilon>)^2>} \qquad . \qquad (30)$$

One observes in Fig. 2 below, that $A_0^1(\vec{q})$ vanishes as $q$ in the long wavelength limit at low temperature $T$ (its slope increasing only slightly when $T$ is doubled.)

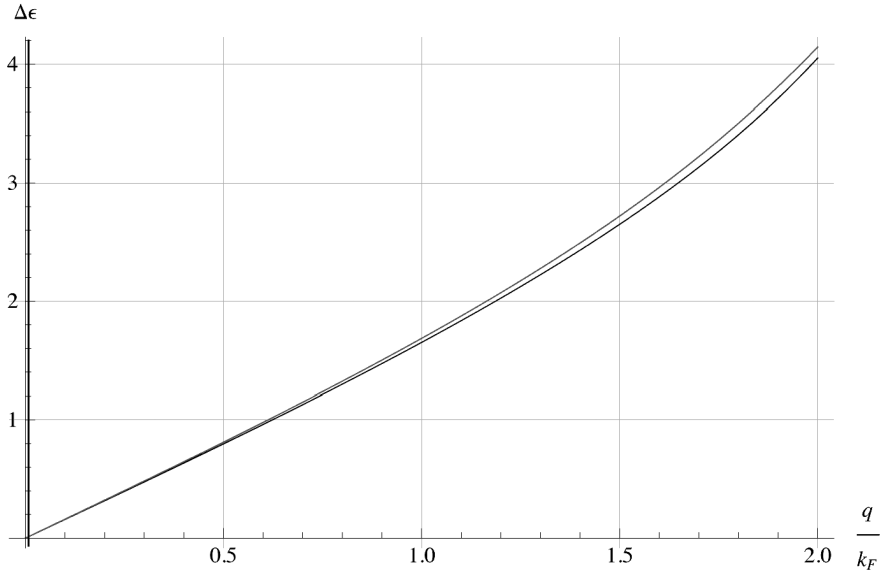

Figure 1. Plot of $A_0{}^0 = <\Delta\varepsilon>$ as calculated in Eqs. (26) or (28), as function of $q/k_F$ in 2D, at $kT= 0.1$ (lower curve) and 0.2 (upper) in the same units as in Fig. 1.

Division of (29) by (30) allows explicit evaluation of the new operator $a_1$ itself:

$$a_1(\vec{q}) = \frac{A_0^1(\vec{q}) a_1(\vec{q})}{A_0^1(\vec{q})}$$

$$= \frac{\dfrac{C(\vec{q})}{\sqrt{|q| Vol}} \sum_k \theta(\vec{k} \cdot \vec{q})(\Delta\varepsilon - <\Delta\varepsilon>) c^{\dagger}(\vec{k} - \vec{q}/2) c(\vec{k} + \vec{q}/2)}{\sqrt{<(\Delta\varepsilon - <\Delta\varepsilon>)^2>}}$$

(31A)

This new normal mode has been explicitly normalized, *by construction*. Its corresponding $\Phi_1$ can be extracted expressed as function of $\Delta\varepsilon$ and its averages. It is seen to exhibit a "radial" node at $\Delta\varepsilon = <\Delta\varepsilon>$,

$$\Phi_1(\vec{k}, \vec{q}) = \frac{(\Delta\varepsilon - <\Delta\varepsilon>)}{\sqrt{<(\Delta\varepsilon - <\Delta\varepsilon>)^2>}}$$

(31B)

**(7) COUNTING NODES.** The above suggests an interesting and most productive way of classifying the $\Phi_j$'s. Inspecting the numerator of Eqs. (31) we observe that the energy of some of the elementary excitations $\Delta\varepsilon = \varepsilon(\vec{k} + \vec{q}/2) - \varepsilon(\vec{k} - \vec{q}/2)$ are lower than average, and some higher, thus proving that $\Phi_1$ possesses *one* extra "*radial* node", *i.e.*, one change of sign in the direction of increasing $\Delta\varepsilon$ ( relative to $\Phi_0 = 1$ which has none.) This radial node is the locus of the *curve* (in 2D) or *surface* (in 3D) defined by solutions of the equation $\Delta\varepsilon - <\Delta\varepsilon> = \varepsilon(\vec{k} + \vec{q}/2) - \varepsilon(\vec{k} - \vec{q}/2) - A_0^0 = 0$ , *i.e.*, $\Phi_1 = 0$.

Next, putting it loosely, $a_2$ has to have 2 non-overlapping radial nodes, as it is constructed orthogonal to $a_0$ and $a_1$, and so on. We conclude that $a_j$ (or, specifically, $\Phi_j$ ) has $j$ radial nodes in $k$-space. In 2D, $\Phi_j$ vanishes on $j$ distinct curves in $k$-space, in 3D on $j$ distinct surfaces, deduced from $\Phi_j$ being a polynomial of degree $j$ in $\Delta\varepsilon$ , having $j$ real roots.

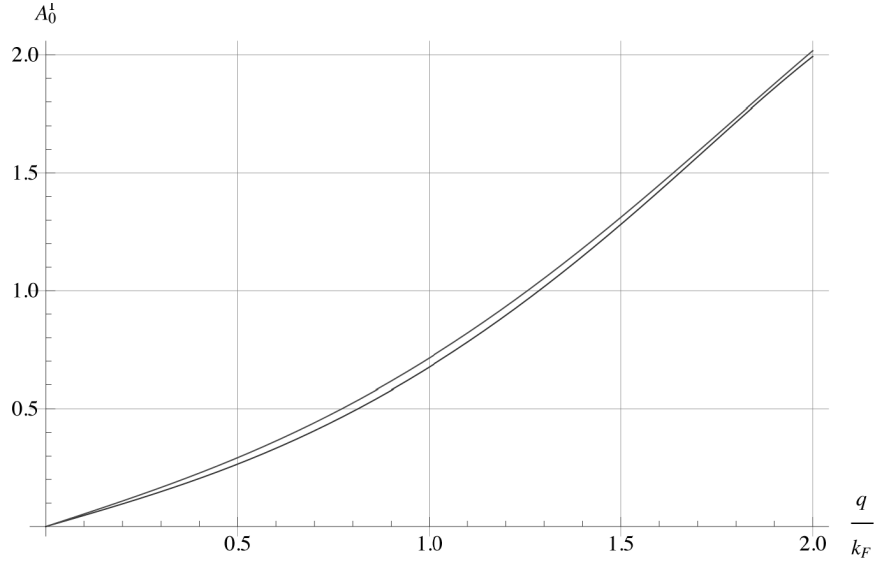

Figure 2. A second coefficient, $A_0^1$ *vs.* $q/k_F$ , at temperatures $kT$=0.1 (lower curve) and 0.2 (upper) in 2D, using the same units as in Fig. 1.

If radial nodes are necessary, then why not transverse nodes? (These are defined as changes in sign in $\Phi_j$ in the direction of *constant* $\Delta\varepsilon = e(\vec{k}+\vec{q}\,/\,2) - e(\vec{k}-\vec{q}\,/\,2)$ .)  Now, given that $\Phi_0$ has neither transverse nor radial nodes, the equations of motion that allow $\Phi_1$ to acquire one *radial* node, $\Phi_2$ two *radial* nodes, etc., will not by themselves ever introduce transverse nodes. No function $\Phi_a$ exhibiting a number $n\neq0$ of *transverse* nodes could ever appear in the commutator equations of motion. So it follows that no corresponding $a_a$ operators can appear in this reformulation of the Hubbard Hamiltonian.

The number of transverse nodes is unaffected by $H_2$ in the Hubbard model, so it is a constant of the motion. That does not mean that the operators with 1 or more transverse nodes can be totally ignored: "completeness" compels us to use *all* possible normal modes in the calculation of certain nontrivial correlation functions and dynamical quantities. But calculation of the all-important ground-state energy (and free energy at finite $T$) requires only knowledge about bosons that are nodeless transversally. This simplification is a unique simplifying feature of the Hubbard model and does not necessarily hold true in other models – especially if we include "exchange" terms.

**(8) SECOND EQUATION OF MOTION AND BEYOND.** We pursue successive commutator bracket "equations of motion", $[a_1(\vec{q}), H_1]$ and beyond, in density (and *spin*) sectors of the Hubbard model, taking advantage of the new "averaging" notation of §6.

First, *in words*: the next iteration produces 3 objects. The first is known: $A_1^0 a_0$, that we rewrite $A_0^1 a_0$.[21] The second is a multiple of $a_1(\vec{q})$, $A_1^1 a_1(\vec{q})$, which defines $A_1^1$. Finally, the remainder, orthogonal to both $a_0$ and $a_1$, is denoted $A_1^2 a_2(\vec{q})$. Here $A_1^1$ and $A_1^2$ are two new coefficients that we need to calculate. All coefficients are, typically, functions of $q$. The operator $a_2(\vec{q})$ is a boson, normalized and orthogonal to $a_0$ and $a_1$; it too is new and remains to be calculated. *Algebraically* we can express these facts so much more succinctly: $[a_1(\vec{q}), H_1] = A_0^1(\vec{q})a_0(\vec{q}) + A_1^1(\vec{q})a_1(\vec{q}) + A_1^2(\vec{q})a_2(\vec{q})$.

$A_0^1(\vec{q})a_1(\vec{q})$ is known from Eqs. (29) and (31), $a_0(\vec{q})$ and $a_1(\vec{q})$ have both been previously defined (*cf*. Eq. (3)) or derived (*cf*. Eq.(31)). We project $A_1^1(\vec{q})$ out using (29) and a nested commutator, $A_1^1(\vec{q}) = \frac{1}{(A_0^1)^2}[[A_0^1 a_1(\vec{q}), H_1], A_0^1 a_1^\dagger(\vec{q})]$. Explicitly,

$$A_1^1(\vec{q}) = \frac{1}{(A_0^1(\vec{q}))^2} < (\varepsilon(\vec{k}+\vec{q}/2) - \varepsilon(\vec{k}-\vec{q}/2) - A_0^0(\vec{q}))^2 (\varepsilon(\vec{k}+\vec{q}/2) - \varepsilon(\vec{k}-\vec{q}/2)) >$$

$$= \frac{<(\Delta\varepsilon - <\Delta\varepsilon>)^2 \cdot (\Delta\varepsilon)>}{<(\Delta\varepsilon - <\Delta\varepsilon>)^2>} = \frac{<\Delta\varepsilon^3 - 2<\Delta\varepsilon^2><\Delta\varepsilon> + <\Delta\varepsilon>^3>}{<\Delta\varepsilon^2 - <\Delta\varepsilon>^2>}$$

(32)

is a real quantity not unlike the corresponding virial coefficients of statistical mechanics. The square of $A_1^2(\vec{q})$ can then be calculated in two steps *before* even knowing $a_2$. First, project out $A_1^2(\vec{q})a_2(\vec{q})$ in the equation of motion,

$$A_1^2(\vec{q})a_2(\vec{q}) = \frac{1}{A_0^1(\vec{q})}\left([A_0^1(\vec{q})a_1(\vec{q}), H_1] - (A_0^1(\vec{q}))^2 a_0(\vec{q}) - A_1^1(\vec{q})A_0^1(\vec{q})a_1(\vec{q})\right)$$

$$= \frac{1}{A_0^1(\vec{q})}\frac{C(\vec{q})}{\sqrt{q \times Vol}}\sum_k \theta(\vec{k}\cdot\vec{q}) \times$$

$$\left\{(\varepsilon(\vec{k}+\vec{q}/2) - \varepsilon(\vec{k}-\vec{q}/2) - A_0^0(\vec{q}))(\varepsilon(\vec{k}+\vec{q}/2) - \varepsilon(\vec{k}-\vec{q}/2) - A_1^1(\vec{q})) - (A_0^1(\vec{q}))^2\right\} \times$$

$$c^\dagger(\vec{k}-\vec{q}/2)c(\vec{k}+\vec{q}/2)$$

(33A)

Then $[A_1^2(\vec{q})a_2(\vec{q}), A_1^2(\vec{q})a_2^\dagger(\vec{q})] = (A_1^2(\vec{q}))^2 = \frac{1}{(A_0^1(\vec{q}))^2} < \left((\Delta\varepsilon - A_0^0)(\Delta\varepsilon - A_1^1) - (A_0^1)^2\right)^2 >$,

the positive root of which yields,

$$A_1^2(\vec{q}) = \frac{1}{A_0^1(\vec{q})}\sqrt{<\left((\Delta\varepsilon - <\Delta\varepsilon>)(\Delta\varepsilon - A_1^1(\vec{q})) - (A_0^1(\vec{q}))^2\right)^2>}$$

(33B)

It is expressed in terms of previously calculated $A_n^m(\vec{q})$ coefficients that can themselves all be expressed in the $<...>$ notation. We plot $A_1^2(\vec{q})$ in Fig. 3 below.

Eqs. (32) and (33) also allow us to spell out the normalized, 2-radial-node boson operator $a_2$ as a sum over elementary excitations, in terms of the known quantities $A_0^0(\vec{q})$, $A_0^1(\vec{q})$ $(= A_1^0(\vec{q}))$, $A_1^1(\vec{q})$ and $A_1^2(\vec{q})$ $(= A_2^1(\vec{q}))$. These coefficients also help to extract the wave function $\Phi_2$ from Eqs. (33),

$$a_2(\vec{q}) = \frac{1}{A_1^2(\vec{q})A_0^1(\vec{q})}\left([A_0^1(\vec{q})a_1(\vec{q}), H_1] - (A_0^1(\vec{q}))^2 a_0(\vec{q}) - A_1^1(\vec{q})A_0^1(\vec{q})a_1(\vec{q})\right).$$

We leave the derivation as an exercise for the reader.

Similarly, the $n^{th}$ turn yields successive $A_n^m = A_m^n$ coefficients with $m=n-1, n$, and $n+1$. Direct calculation shows all coefficients to be real and unique. These generalized equation of motion, for arbitrary $n$, are:

$$[a_n(\vec{q}), H_1] = A_{n-1}^n(\vec{q})a_{n-1}(\vec{q}) + A_n^n(\vec{q})a_n(\vec{q}) + A_n^{n+1}a_{n+1}(\vec{q}), \tag{34A}$$

where $A_0^{-1} = A_{-1}^0 \equiv 0$ serves as the initial (boundary) condition. The conjugate equations are,

$$[H_1, a_n^\dagger(-\vec{q})] = A_{n-1}^n(\vec{q})a_{n-1}^\dagger(-\vec{q}) + A_n^n(\vec{q})a_n^\dagger(-\vec{q}) + A_n^{n+1}(\vec{q})a_{n+1}^\dagger(-\vec{q}). \tag{34B}$$

The coefficients are spatially symmetric $A_n^m(\pm\vec{q}) = A_m^n(\pm\vec{q}) = A_m^n(\mp\vec{q})$ and, in the Hubbard model, numerically the same for the spin-density bosons as in the charge-density sectors.

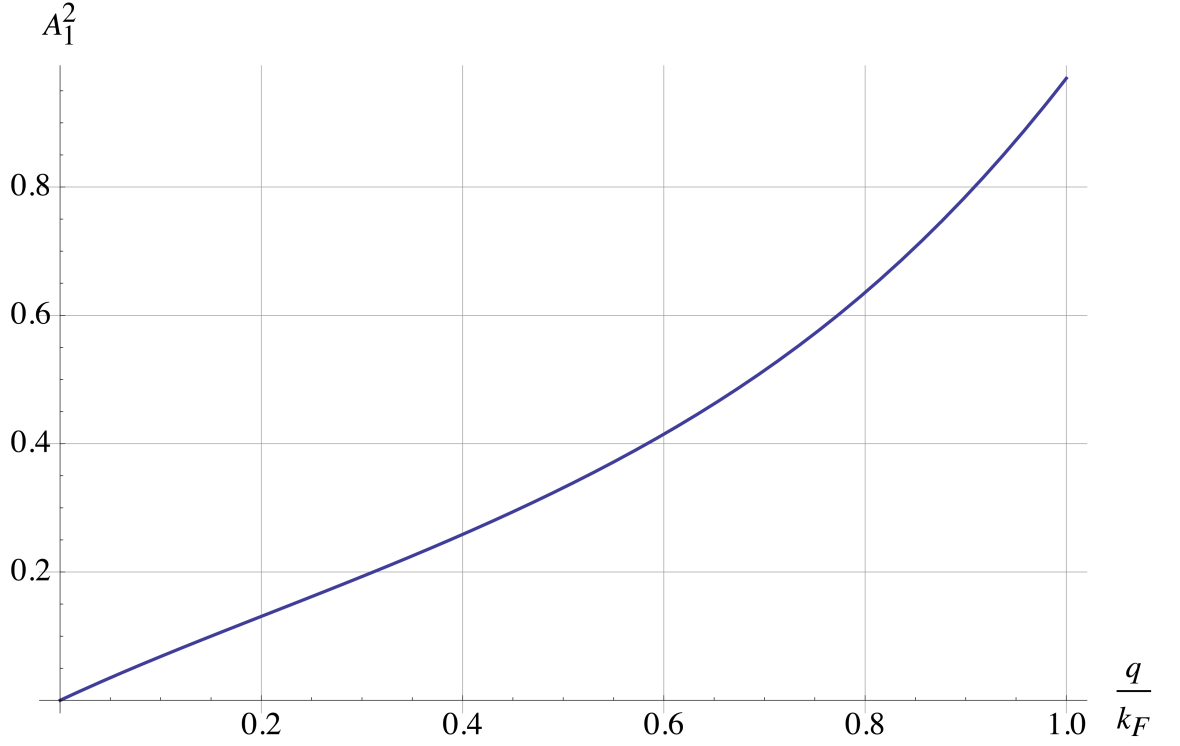

Figure 3. Coefficient $A_1^2(\bar{q})$ as a function of $q/k_F$ in 2D, at $kT$=0.1, in same units as in the preceding figures.

**(9) THE HARMONIC STRING.** Consider the following quadratic form expressed directly in *boson* operators constructed with the aid of the coefficients calculated above,

$$\hat{A}(\bar{q}) = \sum_{n=0}^{\infty} A_n^n(\bar{q})(a_n^\dagger(\bar{q})a_n(\bar{q}) + a_n^\dagger(-\bar{q})a_n(-\bar{q})) +$$
$$\sum_{n=0}^{\infty} \left( A_n^{n+1}(\bar{q})\left(a_n^\dagger(\bar{q})a_{n+1}(\bar{q}) + a_n^\dagger(-\bar{q})a_{n+1}(-\bar{q})\right) + H.c. \right) \quad (35)$$

This defines the Hamiltonian of an "*harmonic string*," $\hat{A}(\bar{q})$. The equations of motion of $a_0(\bar{q})$, $a_1(\bar{q})$, ..., $a_n(\bar{q})$ $w.r.$ to this Hamiltonian are:

$$[a_n(\vec{q}), \hat{A}(\vec{q})] = A_{n-1}^n(\vec{q})a_{n-1}(\vec{q}) + A_n^n(\vec{q})a_n(\vec{q}) + A_n^{n+1}a_{n+1}(\vec{q}) \quad . \tag{36}$$

$H_1$ is interchangeable with $\hat{A}(\vec{q})$ as the *rhs* is visibly identical with Eqs. (34A). (Similarly, equations of the Hermitean conjugate operators are identical with (34B).)

We established previously that the equations of motion do not change the initial number of angular nodes – as opposed to radial nodes. This number is therefore a "constant of the motion." The transversally nodeless bosons with which we have been concerned could be denoted "*s*-waves." Similar sets of equations for bosons with *l*=1 transverse node (*p*-waves) and *n*=0,1,2,... radial nodes, those with *l*=2 angular nodes (*d*-wave), etc., all generate their own sets of *A* coefficients. Numerically, these differ from each other and from those calculated for the *s*-waves given that the $A_n^m(l,\vec{q}) = A_m^n(l,\vec{q})*$ depend on *l* as well as on *q*. Fortunately, in this work, we only have to deal with *l*=0.

When later we add in all the additional contributions from $H_2$ the result will be a set of exactly diagonalizable quadratic forms in boson operators, one in each sector. But before confronting this new task, let us examine all channels, both obvious and novel.

**(10) COOPERON STRINGS.** The equations of motion of cooperons are similar to the preceding but numerically their solutions, coefficients that we shall label *B*, will differ.

Start with the (+) cooperons of Eq. (21),

$$\chi_{j,+}(\vec{q}) = \frac{D_+(\vec{q})}{\sqrt{Vol}} \sum_{\vec{k}} \theta\left(\varepsilon(\vec{k}+\vec{q}/2) + \varepsilon(-\vec{k}+\vec{q}/2)\right) \Psi_{j,+}(\vec{q}\,|\,\vec{k}) c_\downarrow(\vec{k}+\vec{q}/2) c_\uparrow(-\vec{k}+\vec{q}/2)$$

The first equation of motion, $[\chi_{0,+}(\vec{q}), H_1] = B_0^0(\vec{q})\chi_{0,+}(\vec{q}) + B_0^1(\vec{q})\chi_{1,+}(\vec{q})$, when written out, yields:

$$[\chi_{0,+}(\vec{q}), H_1] = \frac{D_+(\vec{q})}{\sqrt{Vol}} \sum_{\vec{k}} \theta\left(\varepsilon(\vec{k}+\vec{q}/2) + \varepsilon(-\vec{k}+\vec{q}/2)\right) \times$$
$$\left(\varepsilon(-\vec{k}+\vec{q}/2) + \varepsilon(\vec{k}+\vec{q}/2)\right) c_\downarrow(\vec{k}+\vec{q}/2) c_\uparrow(-\vec{k}+\vec{q}/2) \tag{37}$$

The first coefficient in the string, $B_0^0(\vec{q})$, is obtained from $[[\chi_{0,+}(\vec{q}), H_1], \chi_{0,+}^\dagger(\vec{q})]$, and is:

$$B_0^0(\vec{q}) = \frac{(D_+)^2}{Vol} \sum_{\vec{k}} \theta\Big(\varepsilon(\vec{k} + \vec{q}/2) + \varepsilon(-\vec{k} + \vec{q}/2)\Big) \times$$
$$\Big(\varepsilon(-\vec{k} + \vec{q}/2) + \varepsilon(\vec{k} + \vec{q}/2)\Big)(1 - f(\vec{k} + \vec{q}/2) - f(-\vec{k} + \vec{q}/2)) \tag{38A}$$

We define a cooperon probability function by analogy with the density- and spin-modes. According to Eq. (38A) it should be,

$$P_{\vec{q},+}^C(\vec{k}) = \frac{(D_+)^2}{Vol} \theta\Big(\varepsilon(\vec{k} + \vec{q}/2) + \varepsilon(-\vec{k} + \vec{q}/2)\Big) \times (1 - f(\vec{k} + \vec{q}/2) - f(-\vec{k} + \vec{q}/2)) . \tag{39A}$$

This $P$ satisfies two important requirements: it is everywhere non-negative and its sum over $k$ is 1 (as required by the normalization of $\chi_{0,+}(\vec{q})$).[17] Making use of this compact notation and writing symbolically, $\varepsilon + \varepsilon' \equiv \varepsilon(\vec{k} + \vec{q}/2) + \varepsilon(-\vec{k} + \vec{q}/2)$, we follow the earlier derivation to obtain the next coefficient. $B_0^0(\vec{q}) = \langle \varepsilon + \varepsilon' \rangle$. Also,

$$(B_0^1(\vec{q}))^2 = [B_0^1(\vec{q}) \chi_{1,+}(\vec{q}), \Big(B_0^1(\vec{q}) \chi_{1,+}(\vec{q})\Big)^\dagger] = \langle (\varepsilon + \varepsilon' - \langle \varepsilon + \varepsilon' \rangle)^2 \rangle, \text{ i.e.,}$$

$$B_0^1(\vec{q}) = \sqrt{\langle (\varepsilon + \varepsilon' - \langle \varepsilon + \varepsilon' \rangle)^2 \rangle} \tag{40}$$

One major difference with the density or spin-modes concerns the cutoff; at large values of the $k$-space variable $\varepsilon + \varepsilon' \equiv \varepsilon(\vec{k} + \vec{q}/2) + \varepsilon(-\vec{k} + \vec{q}/2)$, the probability density (39) tends to a constant, so $P$ cannot be normalized without a cutoff at some $k = q_0$ that is either physically motivated (say, an upper band edge) or is arbitrary.

There is, however, no such conundrum concerning the (-) branch of the cooperons, as $k_F$ provides a natural cutoff. The corresponding probability function is,

$$P_{\vec{q},-}^C(\vec{k}) = \frac{(D_-)^2}{Vol} \theta\Big(-\varepsilon(\vec{k} + \vec{q}/2) - \varepsilon(-\vec{k} + \vec{q}/2)\Big) \times (f(\vec{k} + \vec{q}/2) + f(-\vec{k} + \vec{q}/2) - 1) \tag{39B}$$

Because the formula for $B_0^0(\vec{q})$ acquires a minus sign when evaluated for the (-) branch,

$$B_0^0(\vec{q}) = -\frac{(D_-)^2}{Vol} \sum_{\vec{k}} \theta\left(-\varepsilon(\vec{k}+\vec{q}/2) - \varepsilon(-\vec{k}+\vec{q}/2)\right) \times$$
$$\left(\varepsilon(-\vec{k}+\vec{q}/2) + \varepsilon(\vec{k}+\vec{q}/2)\right)(f(\vec{k}+\vec{q}/2) + f(-\vec{k}+\vec{q}/2) - 1)$$

(38B)

(compactly written as, $B_0^0(\vec{q}) = \ <-(\varepsilon + \varepsilon')>$) , it remains resolutely positive. The formula for $B_0^1(\vec{q})$ remains formally identical to (40) except that averages must be taken $w.r.$ to probabilities (39B) (and not (39A).) The plots in Fig. 4 show $B_0^1(\vec{q})$ for this (-) species to have an approximately parabolic dependence on $q$.

Further iterations proceed in the same manner for the other species.

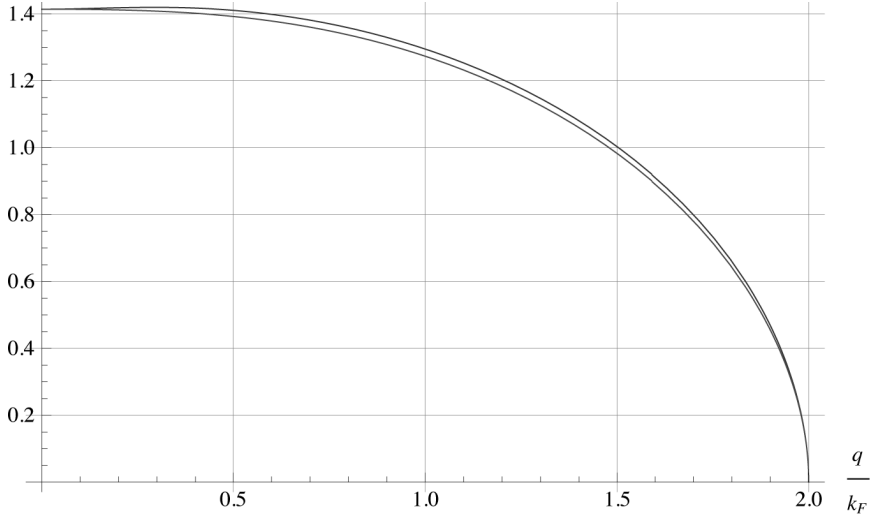

Figure 4. Coefficient $B_0^1(\vec{q})$ $vs.$ $q/k_F$ for the hole-like cooperons in 2D, at two

temperatures, $kT$=0.1 (upper curve) and 0.2, in the same units as previously.

**(11) COMPLETION OF THE STRINGS BY $H_2$.** Ultimately we need to replace $H_1$ by the total Hamiltonian $H = H_1 + H_2$ in the string equations, modifying them in two ways.

Each string devolves from initial operators. The initial operators ($a_0$, $x_0$, etc.) are chosen to be those found in the interaction Hamiltonian $H_2$. This particular choice is what allows the interaction to affect the coefficients minimally, at the zero$^{th}$ sites of each string only. As an example, consider the modifications caused by $H_{2,dir}$ of Eq. (14) in §4, to the equations of motion in the density channel. We recall that $H_{2,dir}$ is:

$$H_{2,dir} = U \sum_{\vec{q},\, q_z > 0} \frac{|q|}{C^2(q)} \left\{ a_0^\dagger(\vec{q}) a_0(\vec{q}) + a_0^\dagger(-\vec{q}) a_0(-\vec{q}) + \left( a_0(\vec{q}) a_0(-\vec{q}) + a_0^\dagger(-\vec{q}) a_0^\dagger(\vec{q}) \right) \right\}$$

The diagonal operator $a_0^\dagger(\vec{q}) a_0(\vec{q}) + a_0^\dagger(-\vec{q}) a_0(-\vec{q})$ clearly adds $U \dfrac{|q|}{C^2(q)}$ to $A_0^0(\vec{q})$ (the coefficient previously calculated in Eq. (28)) without modifying the others.

The nondiagonal operators $a_0(\vec{q}) a_0(-\vec{q}) + a_0^\dagger(-\vec{q}) a_0^\dagger(\vec{q})$ are nontrivial in that they connect strings that were formerly disjoint. Replacing two separate Eqs. (25), pairs of strings merge at $j=0$, where they conjoin operators from both $\pm q$ into a single $\infty$ string as shown below. By momentum conservation the string that commences with $a_0(\vec{q})$ can connect only to the string that commences with $a_0^\dagger(-\vec{q})$. That said, the first of two equations of motion, replacing Eqs. (25), is:

$$[a_0(\vec{q}), H] = \left( A_0^0(\vec{q}) + U \frac{|q|}{C^2(q)} \right) a_0(\vec{q}) + A_0^1(\vec{q}) a_1(\vec{q}) + \left( \frac{U|q|}{C^2(q)} \right) a_0^\dagger(-\vec{q}) \tag{41A}$$

and the other,

$$[H, a_0^\dagger(-\vec{q})] = \left( A_0^0(\vec{q}) + U \frac{|q|}{C^2(q)} \right) a_0^\dagger(-\vec{q}) + A_0^1(\vec{q}) a_1^\dagger(-\vec{q}) + \left( \frac{U|q|}{C^2(q)} \right) a_0(\vec{q}) \tag{41B}$$

Subsequent iterations connecting sites 1 to 2 etc. are unaffected; they remain independent of $U$ and symbolically (and numerically) the same as in (25). We see that strings of annihilation operators at $\vec{q}$ connect to strings of creation operators at $-\vec{q}$ at the origin, $j=0$, but nowhere else.

The preceding suggests renumbering the strings of creation operators emanating from (41B), changing subscript 1 to $-1$ (*i.e.*, renaming $a_1^\dagger(-\vec{q}) \to a_{-1}^\dagger(-\vec{q})$) and, for $j > 1$, relabeling subscripts $j$ as $-j$ (*i.e.*, rename $a_j^\dagger(-\vec{q})$ as $a_{-j}^\dagger(-\vec{q})$) in subsequent iterations. In this way we have mapped the two semi-infinite strings onto *a single*, *infinite*, string, numbered from $-\infty < j < +\infty$. One simplification of this picture is that *all* the effects of the two-body interactions $H_{2,dir}$ become contained on the single, central, site, $j=0$.

Instead of $\hat{A}(\vec{q})$, the corresponding *density* pseudo-Hamiltonian at $\vec{q}$ is an Hermitean quadratic form in the density operators that reads as follows:

$$\hat{H}_{dens}(\vec{q}) = \sum_{n=0}^\infty (A_n^n(\vec{q}) + \delta_{n,0}\frac{U\,|\,q\,|}{C^2(q)}) a_n^\dagger(\vec{q}) a_n(\vec{q}) + \sum_{n=0}^\infty \Big(A_n^{n+1}(\vec{q}) a_n^\dagger(\vec{q}) a_{n+1}(\vec{q}) + H.c.\Big)$$
$$+ \sum_{n=0}^\infty (A_n^n(\vec{q}) + \delta_{n,0}\frac{U\,|\,q\,|}{C^2(q)}) a_{-n}^\dagger(-\vec{q}) a_{-n}(-\vec{q}) + \sum_{n=0}^\infty \Big(A_n^{n+1}(\vec{q}) a_{-n}^\dagger(-\vec{q}) a_{-n-1}(-\vec{q}) + H.c.\Big)$$
$$+ \frac{U\,|\,q\,|}{C^2(q)} \Big(a_0(\vec{q}) a_0(-\vec{q}) + a_0^\dagger(-\vec{q}) a_0^\dagger(\vec{q})\Big) \tag{42}$$

The $\vec{q}$ occupy a single hemisphere (the one with $q_z > 0$ is chosen,) as the expression explicitly includes both $\vec{q}$ and $-\vec{q}$. The corresponding *spin* pseudo-Hamiltonian is,

$$\hat{H}_\sigma(\vec{q}) = \sum_{n=0}^\infty (A_n^n(\vec{q}) - \delta_{n,0}\frac{U\,|\,q\,|}{C^2(q)}) \vec{\sigma}_n^\dagger(\vec{q}) \cdot \vec{\sigma}_n(\vec{q}) + \sum_{n=0}^\infty \Big(A_n^{n+1}(\vec{q}) \vec{\sigma}_n^\dagger(\vec{q}) \cdot \vec{\sigma}_{n+1}(\vec{q}) + H.c.\Big)$$
$$+ \sum_{n=0}^\infty (A_n^n(\vec{q}) - \delta_{n,0}\frac{U\,|\,q\,|}{C^2(q)}) \vec{\sigma}_{-n}^\dagger(-\vec{q}) \cdot \vec{\sigma}_{-n}(-\vec{q}) + \sum_{n=0}^\infty \Big(A_n^{n+1}(\vec{q}) \vec{\sigma}_{-n}^\dagger(-\vec{q}) \cdot \vec{\sigma}_{-n-1}(-\vec{q}) + H.c.\Big)$$
$$- \frac{U\,|\,q\,|}{C^2(q)} \Big(\vec{\sigma}_0(\vec{q}) \cdot \vec{\sigma}_0(-\vec{q}) + \vec{\sigma}_0^\dagger(-\vec{q}) \cdot \vec{\sigma}_0^\dagger(\vec{q})\Big) \tag{43}$$

Recall that the $A$ coefficients of Eq. (43) are numerically identical to those in Eq. (42). The only difference between the two equations is an explicit reversal in the sign of the interaction, $U \leftrightarrow -U$. Cooperon strings present a separate challenge, discussed later.

**(12) THE SOLUTIONS: GENERALITIES.** We have the equations. How do we solve them? At finite $q$ it may not be practical to diagonalize expressions such as (42) or (43); or it may be unwieldy, as this requires knowing all the coefficients $A_n^m$ as functions of $q$, most notably in the asymptotic regions $n \to \pm\infty$. So let us take the time to examine other possibilities. The most obvious is to express exact bosonic raising/lowering operators of the quadratic form as linear combinations of the bare operators. For example, define:

$$\zeta(\vec{q}) = \sum_{j=0}^{\infty} \{F_j z_j(\vec{q}) + G_j z_{-j}^{\dagger}(-\vec{q})\} , \qquad (44)$$

presumed to be a lowering operator of $\hat{H}_z$. For this to hold, $\zeta$ has to satisfy an equation with closure: $[\zeta(\vec{q}), \hat{H}_z(\vec{q})] = \lambda\zeta(\vec{q})$ with $\lambda > 0$. This automatically makes it a lowering operator of $\hat{H}_\sigma(\vec{q}) = \hat{H}_x + \hat{H}_y + \hat{H}_z$ (it commutes with $\hat{H}_x + \hat{H}_y$) and therefore, of the full $H$. Assuming we were able to solve this commutator equation for all its $N \to \infty$ roots, we could use them to transform $\hat{H}_z(\vec{q})$ into diagonal form:

$$\hat{H}_z(\vec{q}) \Rightarrow \sum_{n=0}^{\infty} \lambda_n \{\zeta_n^{\dagger}(\vec{q})\zeta_n(\vec{q}) + \zeta_n^{\dagger}(-\vec{q})\zeta_n(-\vec{q}) + K_n\} \qquad (45)$$

in which $\lambda_n K_n$ are individual contributions to the zero-point energy, the $\lambda_n$ being the eigenvalues (chosen as positive roots in the event of ambiguity).[22] Other Hamiltonians, $\hat{H}_x$ and $\hat{H}_y$, have identical spectra. But how do we calculate these $\lambda_n$? It is frequently possible to transform solutions to equations of motion, such as the two point recursion formulas we developed here, as roots of continued fractions although ultimately this procedure again requires knowledge of all the $A_n^m$'s.

If, at small $q$, we discarded $A_0^1$ and all successive coefficients, the surviving strings would have zero length and are solved trivially. Then each string reduces to the original Luttinger model – except that, unlike the original, it renders service in two or three dimensions. To the extent that the discarded coefficients vanish at small $q$ (we see evidence of this in the Figures) this truncation allows calculation of some asymptotic quantities, such as long-range correlations, etc.

Eigenstates that we seek come in two varieties: (1) "bound-state" solutions that one can characterize as "magnons", "plasmons", or whatever, whose energies $\lambda(q)$ lie outside the continuum of unperturbed energies $\{\Delta\varepsilon\}$ ("outside" can mean either above or below the continuum, but always above 0, as previously remarked), or (2) as "scattering-state" solutions, the energies of which interlace the continuum. Scattering states can be understood as elementary excitations that are perturbed by a cloud of interactions.

The best method for the bound states is also one that is easily implemented. Generalizing (44), it regards $\zeta(\vec{q})$ as the sum of two arbitrary functions of the type,

$$\zeta(\vec{q}) = \zeta_1(\vec{q}) + \zeta_2^\dagger(-\vec{q}) =$$

$$\frac{C(\vec{q})}{\sqrt{2 \times q \times Vol}} \sum_k \Big\{ \theta(\vec{k}\cdot\vec{q})\Phi_1(\vec{k})\Big(c_\uparrow^\dagger(\vec{k}-\vec{q}/2)c_\uparrow(\vec{k}+\vec{q}/2) - c_\downarrow^\dagger(\vec{k}-\vec{q}/2)c_\downarrow(\vec{k}+\vec{q}/2)\Big)$$

$$+\theta(-\vec{k}\cdot\vec{q})\Phi_2(\vec{k})\Big(c_\uparrow^\dagger(\vec{k}-\vec{q}/2)c_\uparrow(\vec{k}+\vec{q}/2) - c_\downarrow^\dagger(\vec{k}-\vec{q}/2)c_\downarrow(\vec{k}+\vec{q}/2)\Big)\Big\}$$

$$(46)$$

The solution of $\Big[[\zeta(\vec{q}),\hat{H}_z(\vec{q})] - \lambda(\vec{q})\zeta(\vec{q}), c_\sigma^\dagger(\vec{k}-\vec{q}/2)c_\sigma(\vec{k}+\vec{q}/2)\Big] = 0$ at each $(k,\sigma)$ yields the two wavefunctions $\Phi_1$ and $\Phi_2$. The corresponding eigenvalue $\lambda(q)$ is computed next, in §13.

**(13) THE SOLUTIONS: DO MAGNONS EXIST?** $H$ consists of two parts: $H_1$ and $H_2$. Let us start with just $H_1$ in the commutator; following this we can add $H_2$ into the mix. The first part of the calculation involves $[\zeta(\vec{q}),\hat{H}_z(\vec{q})] - \lambda(\vec{q})\zeta(\vec{q})$, *i.e.*,

$$\frac{C(\vec{q})}{\sqrt{2 \times q \times Vol}} \sum_k \left\{ \theta(\vec{k} \cdot \vec{q}) \Phi_1(\vec{k}) \times \right.$$

$$\left( (2\vec{k} \cdot \vec{q}) - \lambda \right) \left( c_\uparrow^\dagger(\vec{k} - \vec{q}/2) c_\uparrow(\vec{k} + \vec{q}/2) - c_\downarrow^\dagger(\vec{k} - \vec{q}/2) c_\downarrow(\vec{k} + \vec{q}/2) \right)$$

$$+ \tag{47A}$$

$$\theta(-\vec{k} \cdot \vec{q}) \Phi_2(\vec{k}) \times$$

$$\left. \left( (2\vec{k} \cdot \vec{q}) - \lambda \right) \left( c_\uparrow^\dagger(\vec{k} - \vec{q}/2) c_\uparrow(\vec{k} + \vec{q}/2) - c_\downarrow^\dagger(\vec{k} - \vec{q}/2) c_\downarrow(\vec{k} + \vec{q}/2) \right) \right\}$$

The second part, $[\zeta(\vec{q}), \hat{H}_{2,z}(\vec{q})]$, is,

$$-U \frac{|q|}{C^2(q)} \left[ \left\{ \zeta_1(\vec{q}) + \zeta_2^\dagger(-\vec{q}) \right\}, \left\{ z_0^\dagger(\vec{q}) z_0(\vec{q}) + z_0^\dagger(-\vec{q}) z_0(-\vec{q}) + \left( z_0(\vec{q}) z_0(-\vec{q}) + H.c. \right) \right\} \right] \tag{47B}$$

Calculation of the nonvanishing contributions to (B) involve such quantities as,

$$[\zeta_1(\vec{q}), z_0^\dagger(\vec{q})] = \frac{C^2(q)}{q \times Vol} \sum_k \theta(\vec{k} \cdot \vec{q}) \Phi_1(\vec{k}) \{ f(\vec{k} - \vec{q}/2) - f(\vec{k} + \vec{q}/2) \} = \; <\Phi_1(\vec{k})>_+ \; \text{and}$$

$$[\zeta_2^\dagger(-\vec{q}), z_0(-\vec{q})] = - <\Phi_2(\vec{k})>_- \; \text{where} <...>_\pm \; \text{indicates that the averaging is over one or}$$

the other of the two nonintersecting ensembles, $\vec{k} \cdot \vec{q} > 0$ for $\Phi_1$ and of $\vec{k} \cdot \vec{q} < 0$ for $\Phi_2$.

(The notation proves its worth in intricate calculations such as this!) Thus, (47B) can be written alternatively as,

$$-U \frac{|q|}{C^2(q)} (<\Phi_1> - <\Phi_2>) \left( z_0^\dagger(-\vec{q}) + z_0(\vec{q}) \right) \tag{47B}$$

In Eq. (47A) the coefficient of the following operator:

$$\frac{C(\vec{q})}{\sqrt{2 \times q \times Vol}} \theta(\vec{k} \cdot \vec{q}) \left( c_\uparrow^\dagger(\vec{k} - \vec{q}/2) c_\uparrow(\vec{k} + \vec{q}/2) - c_\downarrow^\dagger(\vec{k} - \vec{q}/2) c_\downarrow(\vec{k} + \vec{q}/2) \right)$$

is $\Phi_1(\vec{k} \cdot \vec{q}) \left( (2\vec{k} \cdot \vec{q}) - \lambda \right)$. In Eq. (47B) the coefficient of the same operator is:

$$-U \frac{|q|}{C^2(q)} (<\Phi_1>_+ - <\Phi_2>_-). \; \text{For the sum of (47A) and (47B) to vanish as required, the}$$

following must hold:

$$\Phi_1(\vec{k}) = U \frac{|q|}{C^2(q)} \frac{(<\Phi_1>_+ - <\Phi_2>_-)}{(2\vec{k} \cdot \vec{q} - \lambda)} \tag{48A}$$

and similarly,

$$\Phi_2(\vec{k}) = U \frac{|q|}{C^2(q)} \frac{\left( <\Phi_1>_+ - <\Phi_2>_- \right)}{\left( -|2\vec{k}\cdot\vec{q}| - \lambda \right)} \tag{48B}$$

Averaging each $\Phi_j$ over its appropriate ($\pm$) ensemble, we obtain two equations in two unknowns:

$$<\Phi_1>_+ = \left( <\Phi_1>_+ - <\Phi_2>_- \right) \frac{U}{Vol} \sum_{\vec{k}} \theta(\vec{k}\cdot\vec{q}) \frac{f(\vec{k}-\vec{q}/2) - f(\vec{k}+\vec{q}/2)}{\left( 2\vec{k}\cdot\vec{q} - \lambda \right)} , \text{ and} \tag{49A}$$

$$<\Phi_2>_- = \left( <\Phi_1>_+ - <\Phi_2>_- \right) \frac{U}{Vol} \sum_{\vec{k}} \theta(-\vec{k}\cdot\vec{q}) \frac{f(\vec{k}+\vec{q}/2) - f(\vec{k}-\vec{q}/2)}{\left( -|2\vec{k}\cdot\vec{q}| - \lambda \right)} \tag{49B}$$

By an obvious symmetry the integral on the *rhs* of (B) can also be rewritten as:

$$\frac{U}{Vol} \sum_{\vec{k}} \theta(\vec{k}\cdot\vec{q}) \frac{f(\vec{k}-\vec{q}/2) - f(\vec{k}+\vec{q}/2)}{\left( -2\vec{k}\cdot\vec{q} - \lambda \right)} \qquad .$$

Thus, for (A) and (B) to have a simultaneous solution, either $\left( <\Phi_1> - <\Phi_2> \right) = 0$ or

$$1 = \frac{4U}{Vol} \vec{q} \cdot \sum_{\vec{k}} \vec{k} \left( \theta(\vec{k}\cdot\vec{q}) \frac{f(\vec{k}-\vec{q}/2) - f(\vec{k}+\vec{q}/2)}{4(\vec{k}\cdot\vec{q})^2 - \lambda(q)^2} \right) \tag{50}$$

Eq. (50) takes the form, $\frac{1}{U} = \mathtt{M}(q, \frac{\lambda(q)}{2q})$. Here, $\mathtt{M}$ the integral in (50) is (in the thermodynamic limit, once again assuming 2D):

$$\mathtt{M}(q, s(q)) = \frac{1}{q\pi^2} \int dk_y \int_0^\infty dk_x k_x \frac{f(\vec{k}-\vec{q}/2) - f(\vec{k}+\vec{q}/2)}{4k_x^2 - s^2(q)} \qquad . \quad \text{The } x\text{-axis has been chosen}$$

along $\vec{q}$ and $s(q) = \frac{\lambda(q)}{q} \approx s(0)$ is the phase velocity of the normal mode – if there is one.

At finite $T$ the pole at $4k_x^2 - s^2(q) = 0$ causes $\mathtt{M}$ to be complex, hence it causes all states (this includes bound states, if any) to have finite lifetimes.

But in this instance we conclude that even at $T$=0, there *are* in fact no bound states. For plotting $\mathtt{M}$ as function of $s$ shows that in the parameter space where it is not complex (that is, in the region $s > 2$, more or less), M is always *negative*. This holds true over the entire range of reasonable values of $q$ and $s(q)$. Therefore M cannot equal $1/U$ which is *positive* in the Hubbard model.

We conclude that in the absence of any external agency establishing long-range order, or in absence of symmetry breaking of *some* sort, the Hubbard model does not – and cannot – spontaneously sustain a stable spectrum of magnons, just as we assumed (provisionally) in §4. But this does not preclude a spectrum of *magnetic scattering states*, the energies of which interlace the continuum, but that cannot be distinguished as individual particles or states.

To introduce these, most briefly, let us select an "initial" operator,

$$\Theta_{\vec{q}}(\vec{k}) \equiv \frac{\theta(\vec{k} \cdot \vec{q})}{\sqrt{2}} \left( c_{\uparrow}^{\dagger}(\vec{k} - \vec{q}/2) c_{\uparrow}(\vec{k} + \vec{q}/2) - c_{\downarrow}^{\dagger}(\vec{k} - \vec{q}/2) c_{\downarrow}(\vec{k} + \vec{q}/2) \right) \qquad (51)$$

(if the reader prefers "state" to operator, she can imagine imagine (51) acting on the ground state) and a corresponding *exact* operator that includes all scattering processes,

$$\Omega_{\vec{q}}(\vec{k}) = \Theta_{\vec{q}}(\vec{k}) + \frac{1}{Vol} \sum_{k' \neq k} L_{\vec{q}}(\vec{k}, \vec{k}') \Theta_{\vec{q}}(\vec{k}') + \frac{1}{Vol} \sum_{k''} M_{\vec{q}}(\vec{k}, \vec{k}'') \Theta_{-\vec{q}}^{\dagger}(\vec{k}'') \qquad (52)$$

The energy eigenvalue of $\Omega_{\vec{q}}(\vec{k})$ is, to O($1/Vol$), the same as the energy of (51), $2\vec{k} \cdot \vec{q}$. So the equations of motion principally serve to calculate the scattering amplitudes $L_{k,k'}$ and $M_{k,k''}$ and the elastic scattering cross-section, $\dfrac{1}{\tau_{\vec{q}}(\vec{k})}$. This last is in turn related to integrals such as $\dfrac{1}{Vol} \text{Im} \left\{ \sum_{k'} L_q(\vec{k}, \vec{k}') \right\}$ that can be calculated in closed form.

**(14) THE SOLUTIONS: ZERO SOUND.** The equations of motion of the particle density modes $a_j$ are, after replacement of $U$ by $-U$ in all the equations, identical to those of the magnons. So where magnons could not be sustained we must now conclude that density-waves *are* coherent. This type of longitudinal normal mode has been called *"zero-sound."* It is the collective mode of a charge-neutral gas with short-ranged two-body repulsions such as in Hubbard's model. In Fig. 5 we plot $s=s(0)$, the speed of sound as a function of $U$ at $q=0$, being the numerical solution of $\dfrac{-1}{U} = \texttt{M}(0, \dfrac{s(0)}{2})$ in 2D.

The figure shows that $s$ is never less than 2, the Fermi velocity in the present units (recall: $k_f=1$, $e_f=1$, $v_f=2$,) and that $s$ is a monotonically increasing function of the coupling constant. By looking at the imaginary part of the integral $\texttt{M}$ at finite temperature $T$, one should find it possible to evaluate the lifetime of the sound wave as function of $U$ and $T$, a lifetime that is – of course– infinite at $T=0$.

In addition, where the energy of sound waves coincides with the continuum, the eigenstates become scattering states – and are treated similarly to those discussed in the preceding section.

**(15) THE SOLUTIONS: ARE THERE ANY COOPERONS ?** By analogy with the preceding paragraphs, we seek to solve the equations of motion of stable cooperon collective modes – if they exist – not by diagonalizing their strings but directly by solving for the envelope wavefunctions of the corresponding collective boson operators. We set the annihilation operator $\chi(\vec{q}) \equiv \left( \chi_+(\vec{q}) + \chi_+^\dagger(-\vec{q}) + \chi_-(\vec{q}) + \chi_-^\dagger(-\vec{q}) \right)$ as some arbitrary linear combination of outer and inner cooperon operators. That is,

$$
\begin{aligned}
\chi(\vec{q}) = & \frac{D_+(\vec{q})}{\sqrt{Vol}} \sum_{\vec{k}} \theta\left( 2\vec{k}^2 + \frac{\vec{q}^2}{2} - 2 \right) \times \\
& \left( \Psi_1(\vec{k}) c_\downarrow(\vec{k} + \vec{q}/2) c_\uparrow(-\vec{k} + \vec{q}/2) + \Psi_2(\vec{k}) c_\uparrow^\dagger(-\vec{k} - \vec{q}/2) c_\downarrow^\dagger(\vec{k} - \vec{q}/2) \right) \\
+ & \\
& \frac{D_-(\vec{q})}{\sqrt{Vol}} \sum_{\vec{k}} \theta\left( 2 - 2\vec{k}^2 - \frac{\vec{q}^2}{2} \right) \times \\
& \left( \Psi_3(\vec{k}) c_\downarrow(\vec{k} + \vec{q}/2) c_\uparrow(-\vec{k} + \vec{q}/2) + \Psi_4(\vec{k}) c_\uparrow^\dagger(-\vec{k} - \vec{q}/2) c_\downarrow^\dagger(\vec{k} - \vec{q}/2) \right)
\end{aligned}
\tag{53}
$$

Then, in order to belong to eigenvalue $\lambda > 0$ , this boson operator must satisfy:

$$[\chi(\bar{q}), H] - \lambda(\bar{q})\chi(\bar{q}) = 0 \qquad . \tag{54}$$

The commutator that involves the interaction is, explicitly:

$$\begin{aligned}
\left[\chi(\bar{q}), H_{2 \cdot \chi}\right] = &+ U\left[\chi(\bar{q}), \left(\frac{\chi_{o,-}(-\bar{q})}{D_-(\bar{q})} + \frac{\chi_{o,+}^\dagger(\bar{q})}{D_+(\bar{q})}\right) \cdot \left(\frac{\chi_{o,-}^\dagger(-\bar{q})}{D_-(\bar{q})} + \frac{\chi_{o,+}(\bar{q})}{D_+(\bar{q})}\right)\right] \\
&+ U\left[\chi(\bar{q}), \left(\frac{\chi_{o,-}(\bar{q})}{D_-(\bar{q})} + \frac{\chi_{o,+}^\dagger(-\bar{q})}{D_+(\bar{q})}\right) \cdot \left(\frac{\chi_{o,-}^\dagger(\bar{q})}{D_-(\bar{q})} + \frac{\chi_{o,+}(-\bar{q})}{D_+(\bar{q})}\right)\right]
\end{aligned} \tag{55}$$

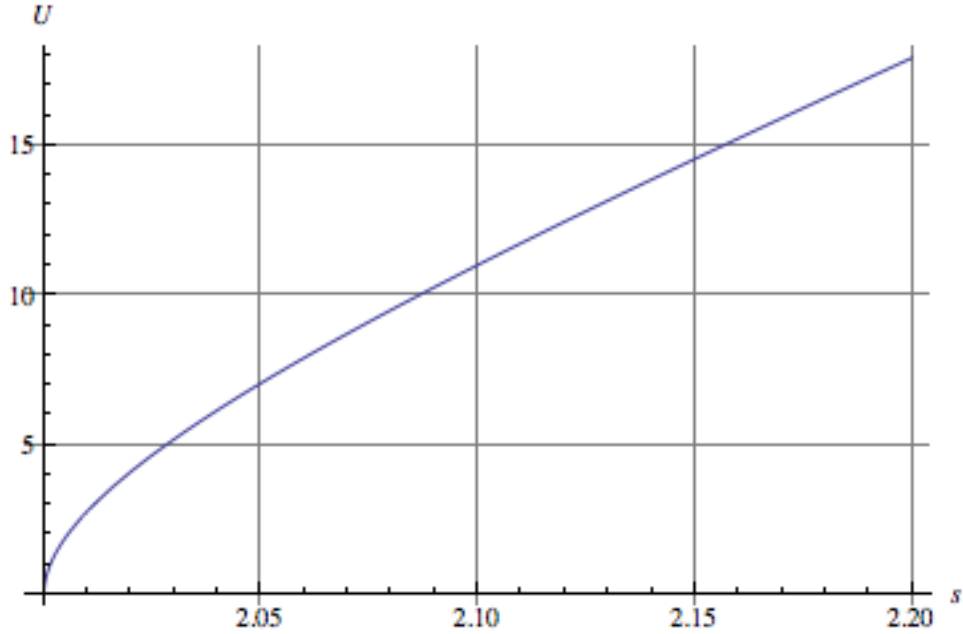

Figure 5. The parameter $U$ (in arbitrary units) *vs.* $s$, the speed of zero-sound in the Hubbard model, at $T$=0, in 2D. (The speed of a particle at the Fermi surface is $v_F$=2. This, in our units and in the absence of interactions, is the limiting value of $s$ at $U$=0.)

The rest of the calculation involves the kinetic energy, that is, $\left[\chi(\bar{q}), H_1\right] - \lambda(\bar{q})\chi(\bar{q})$, where:

$$\left[\chi(\vec{q}), H_1\right] = \frac{D_+(\vec{q})}{\sqrt{Vol}} \sum_{\vec{k}} \theta\left(\varepsilon(\vec{k}+\vec{q}/2) + \varepsilon(-\vec{k}+\vec{q}/2)\right)\left(\varepsilon(\vec{k}+\vec{q}/2) + \varepsilon(-\vec{k}+\vec{q}/2)\right) \times$$

$$\left(\Psi_1(\vec{k})c_\downarrow(\vec{k}+\vec{q}/2)c_\uparrow(-\vec{k}+\vec{q}/2) - \Psi_2(\vec{k})c_\uparrow^\dagger(-\vec{k}-\vec{q}/2)c_\downarrow^\dagger(\vec{k}-\vec{q}/2)\right)$$

$$+$$

$$\frac{D_-(\vec{q})}{\sqrt{Vol}} \sum_{\vec{k}} \theta\left(-\varepsilon(\vec{k}+\vec{q}/2) - \varepsilon(-\vec{k}+\vec{q}/2)\right)\left(\varepsilon(\vec{k}+\vec{q}/2) + \varepsilon(-\vec{k}+\vec{q}/2)\right) \times$$

$$\left(\Psi_3(\vec{k})c_\downarrow(\vec{k}+\vec{q}/2)c_\uparrow(-\vec{k}+\vec{q}/2) - \Psi_4(\vec{k})c_\uparrow^\dagger(-\vec{k}-\vec{q}/2)c_\downarrow^\dagger(\vec{k}-\vec{q}/2)\right) \tag{56}$$

After some algebra, the solutions to Eq.(54) lead to *two* dichotomies:

$$\text{Either } \left(\frac{<\Psi_1>_+}{D_+(\vec{q})} - \frac{<\Psi_3>_-}{D_-(\vec{q})}\right) = 0 \text{ or } \frac{1}{U} = \left(\frac{S_-(-\lambda(q))}{D_-} - \frac{S_+(\lambda(q))}{D_+}\right), \tag{57A}$$

and/or, $\left(\dfrac{<\Psi_2>_+}{D_+(\vec{q})} - \dfrac{<\Psi_4>_-}{D_-(\vec{q})}\right) = 0$ or $\dfrac{1}{U} = \left(\dfrac{S_+(-\lambda(q))}{D_+} - \dfrac{S_-(\lambda(q))}{D_-}\right)$ (57B)

where $S_-(\lambda(q)) = \dfrac{1}{Vol}\sum_{\vec{k}} \dfrac{\theta(2-2k^2-\frac{q^2}{2})(f(\vec{k}+\vec{q}/2) + f(\vec{k}-\vec{q}/2) - 1)}{2-2k^2-\frac{q^2}{2}-\lambda(q)}$

and $S_+(\lambda(q)) = \dfrac{1}{Vol}\sum_{\vec{k}} \dfrac{\theta(2k^2+\frac{q^2}{2}-2)(1-f(\vec{k}+\vec{q}/2) - f(\vec{k}-\vec{q}/2))}{2k^2+\frac{q^2}{2}-2-\lambda(q)}$ ,

assuming, as always, that $\varepsilon(k)=k^2-1$ for $k>1$ and $1-k^2$ for $k<1$. Here, as was previously the case for the magnons, denominators that change sign can cause the integrals $S_\pm(\lambda)$ to be complex for $\lambda > 0$ (by contrast, the integrals $S_\pm(-\lambda)$ are both real and $> 0$).[23] It is possible to solve Eq. (54) for the real and imaginary parts of $\lambda(q)$ but it would be wrong, because these cooperons have a *finite* lifetime even at $T=0$ and cannot qualify as stationary states. Indeed, such scattering states need to be analyzed using scattering theory instead, such as was discussed in §13.

**(16) COLLECTIVE MODES IN THE PRESENCE OF COULOMB OR OTHER FORCES.** For arbitrary potentials the direct interaction terms – that is, those not related either to exchange or pairing phenomena – once again involve just the density operators $\varrho$. These are proportional to what we denoted the particle-density operators $a_0$. Replacing the Hubbard model's Eq. (14) we find, quite more generally,

$$H_{2,dir} = \sum_{\bar{q}, q_z > 0} \frac{V(\bar{q}) | q |}{C^2(q)} \left\{ a_0^{\dagger}(\bar{q}) a_0(\bar{q}) + a_0^{\dagger}(-\bar{q}) a_0(-\bar{q}) + \left( a_0(\bar{q}) a_0(-\bar{q}) + H.c. \right) \right\} \ .$$

The associated density string is therefore the same as before, *except* that the site at $j$=0 hosts a coupling constant $V(q)$ at each $q$ instead of a constant $U$.

By contrast, however, the *exchange* and *pairing* correlations are profoundly affected by changes in the structure of the two-body potentials. For *arbitrary* potentials the exchange and pairing terms can no longer be understood solely using strings that have just nearest-neighbor connections. The reader may wish to mull on it, and we shall return to this topic in a separate publication, but here let us just appreciate that it is the structureless nature of the interactions in momentum space that allows us to solve the many-body problem straightforwardly and exactly in *all* the channels, for the Hubbard model.

**17. GREEN FUNCTIONS AND THE FREE ENERGY.** From elementary thermodynamics we know that the free energy $F$ is an important quantity to know, as it reveals many of the collective properties of a system: its internal energy $E = \frac{\partial(\beta F)}{\partial \beta}$ (where $\beta = 1 / kT$) and specific heat $c_v(T) = \frac{1}{Vol} \frac{dE}{dT}$, the system's entropy $S(T) = -\frac{\partial F}{\partial T}$ and so on. It is generally advantageous to separate $F$ into two parts: the ideal-gas contribution $F_0$ of a noninteracting system having the same number of particles in the same volume and the additional contributions attributed to the interactions among the particles. These number six here: one attributable to density-density fluctuations, three attributable to spin-spin fluctuations, and one in each of two distinct cooperon channels. In the absence of long-range order tying some of them together, we found all these channels to be independent of one another.

With $\{\ldots\}$ indicating the nature of the bosons that are involved in creating each expression, these individual contributions are: $\Delta F_1 = \Delta F(U,T,\{a(q)\})$,

$\Delta F_2 = \Delta F(U,T,\{\sigma_x(q)\})$, $\Delta F_3 = \Delta F(U,T,\{\sigma_y(q)\})$, $\Delta F_4 = \Delta F(U,T,\{\sigma_x(q)\})$,

$\Delta F_5 = \Delta F(U,T,\{\chi_-(q)\})$ and $\Delta F_6 = \Delta F(U,T,\{\chi_+(q)\})$.

They are not all independent. As we remarked earlier, $\Delta F_2 = \Delta F_3 = \Delta F_4$ and $\Delta F_2(U,T) = \Delta F_1(-U,T)$. Thus $F = F_0 + \Delta F_1(U,T) + 3\Delta F_2(U,T) + \Delta F_5(U,T) + \Delta F_6(U,T)$, where

$$F_0 = -2kT\sum_{\bar{k}} \log\left(1 + e^{-\beta\varepsilon(\bar{k})}\right) \quad . \tag{58}$$

For the remainder let us use Feynman's theorem and integrate on the coupling constant $U$. For example, using $<op>_u$ to stand for "op" averaged over states appropriate to a coupling constant $u$, we can express the density contribution to the *free energy*, as,

$$\Delta F_1 =$$
$$\sum_{\bar{q},\, q_z > 0} |q| \int_0^U \frac{du}{C^2(q)} < \left\{a_0^\dagger(\bar{q})a_0(\bar{q}) + a_0^\dagger(-\bar{q})a_0(-\bar{q}) + \left(a_0(\bar{q})a_0(-\bar{q}) + a_0^\dagger(-\bar{q})a_0^\dagger(\bar{q})\right)\right\} >_u \tag{59}$$

to be distinguished from the corresponding interaction *energy*, which is,

$$E_1 = U\sum_{\bar{q},\, q_z > 0} \frac{|q|}{C^2(q)} < \left\{a_0^\dagger(\bar{q})a_0(\bar{q}) + a_0^\dagger(-\bar{q})a_0(-\bar{q}) + \left(a_0(\bar{q})a_0(-\bar{q}) + a_0^\dagger(-\bar{q})a_0^\dagger(\bar{q})\right)\right\} >_U$$

The expectation values in either case are obtained with the aid of Green function techniques in many-body physics. The present example of bosons interacting *via* harmonic forces is easily and satisfactorily dealt with by this method.[24]

For example,

$$
\begin{aligned}
&< a_0^\dagger(\vec{q}) a_0(\vec{q}) >_u = \\
&\frac{1}{2\pi i} \int d\omega \frac{1}{e^{\beta\omega} - 1} \left\{ << a_0(\vec{q}) \mid a_0^\dagger(\vec{q}) >>_u (\omega - i\delta) - << a_0(\vec{q}) \mid a_0^\dagger(\vec{q}) >>_u (\omega + i\delta) \right\}
\end{aligned}
\tag{60}
$$

where $<< a_0(\vec{q}) \mid a_0^\dagger(\vec{q}) >>_u (\omega - i\delta)$ stands for the appropriate "retarded" Green function, evaluated at coupling constant $u$ and frequency $\omega - i\delta$, where $\delta$ is infinitesimal. It can be calculated *exactly* from the equations of motion with the aid of the Hamiltonian quadratic in the harmonic operators.

There is one obvious exception: $C^2(q)$ in the denominator of (59) requires detailed knowledge of $f(k)$ at $U$ and $T$, as may some other correlation functions.

**(18) THE ONE-BODY DISTRIBUTION FUNCTION $f(k)$.** The free energy of the many-fermion system we have been studying is an extensive (that is, proportional to the volume) function, $F(U, T, \{\varepsilon(k)\})$. In the various calculations we have chosen $\varepsilon(k) = k^2 - 1$, with the Fermi wave vector being at $k_F = 1$. To know the *interaction energy*, a sum of correlation functions, we need to compute the *derivative* $U \frac{\partial F}{\partial U}$ (itself a function of $U, T$, and the set of one-particle energies $\{\varepsilon(k)\}$.) Similarly, to know the one-body correlation function $f(\vec{k})$ it is advantageous to compute the *functional derivative* of the free energy, *i.e.*, $< \tilde{n}(\vec{k}) > = f(\varepsilon(\vec{k})) = \frac{1}{2} \frac{\delta F}{\delta \varepsilon(\vec{k})}$ (using the factor ½ to take spin degeneracy into account and treating the individual $\varepsilon(k)$ as a coupling constant.)

This calculation may not always be practical although one could take advantage of circular (or spherical) symmetry to evaluate a simpler derivative,

$$\frac{1}{Vol}\frac{\partial F}{\partial \varepsilon} = \frac{1}{Vol}\sum_{\vec{k}}\sum_{\sigma=\pm}\delta(\varepsilon-\varepsilon(\vec{k}))f(\varepsilon(\vec{k})) = 2\rho(\varepsilon)f(\varepsilon) \qquad (61)$$

where $\varrho(\varepsilon)$ is the (known) density of states. $F = F_0 + \Delta F$ and $f = f_0 + \Delta f$, with $f_0 = \dfrac{1}{e^{\beta\varepsilon}+1}$ being the ordinary Fermi function. We can use the formalism of Eq. (61) to find $\Delta f$, the change in the one-body distribution function due to the interactions, as

$$\Delta f(\varepsilon) = \frac{1}{Vol}\frac{1}{2\rho(\varepsilon)}\frac{\partial \Delta F}{\partial \varepsilon} \qquad (62)$$

The ensuing calculation might appear to be a formidable task. Actually, in the case of the two-dimensional Hubbard model, which we address specifically in the text that follows, the relevant calculations were already performed – some two decades ago – by Hua Chen[25] while studying the normal phase of high-$T_c$ superconductors within the random-phase approximation (*RPA*). Calculations based on the present string theory would be similar if not identical, therefore in what follows we just quote the original results.[25]

Defining the drop in the distribution function at the Fermi level as $Z_F$ ($0 < Z_F < 1$) one finds $Z_F$ to be 1 at $T=0$ when $U=0$, then to decrease smoothly as $U$ is increased until at $U=U_c$ it vanishes. This critical point was calculated at $T=0$ to be $U_c=18$ in units $e_F=1$ and decreases with increasing $T$ (although this last issue was not investigated.)

For $U < U_c$, when the fermions are in a Landau Fermi liquid phase,[24] the inverse lifetime $1/\tau$ of a particle near the Fermi surface, at $|\varepsilon|$, is $1/\tau \propto \varepsilon^2$, *i.e.*, $1/\tau_F \propto T^2$. As $U$ enters the "marginal" or "quantum" Fermi liquid phase at $U \geq U_c$, $f(\varepsilon)$ becomes continuous. Its dependence on $\varepsilon$ near the Fermi surface remains singular, approximately[26]

$$f(\varepsilon) = \frac{1}{2}(1 \pm A \mid \varepsilon \mid^{\delta}),$$ with inverse lifetime $1/\tau \propto \mid \varepsilon \mid$, *i.e.*, $1/\tau_F \propto T$ (at all $\delta \geq 0$). This

so-called "marginal Fermi gas" behavior has in fact been identified in the normal phase of optimally doped two-dimensional high-$T_c$ superconductors.[27]

The one-body distribution function $f(\vec{k})$ satisfies an integral equation in which a kernel $J$ receives contribution from all 6 channels discussed previously. However, only the magnon channels contribute substantially at the Fermi surface and they are – almost entirely – responsible for the decrease in the discontinuity ($Z_F$) at the Fermi level from 1 (at $U = 0$) to 0 (at $U = U_c$.) The relevant equation is:

$$f(\vec{k}) = f_0(\vec{k}) + \frac{3}{2\pi Vol} \operatorname{Re}\left\{ \sum_{\vec{k}'} \int_0^\infty dv J(iv, |\vec{k} - \vec{k}'|) \frac{f(\vec{k}) - f(\vec{k}')}{(iv + e_k - e_{k'})^2} \right\} \tag{63}$$

The essential contribution to $J$ in Eq. (63) is $J(\omega, |\vec{q}|) = \dfrac{U^2 \Pi(\omega, |\vec{q}|)}{1 + U \Pi(\omega, |\vec{q}|)}$, in which the

Lindhard function or "*polarization part*," is $\Pi(\omega, |\vec{q}|) = \dfrac{1}{Vol} \sum_k \dfrac{f(\vec{k} + \vec{q}/2) - f(\vec{k} - \vec{q}/2)}{\omega + e_{\vec{k} + \vec{q}/2} - e_{\vec{k} - \vec{q}/2}}$ .

**19. CONCLUSION.** As we have just witnessed, a "string" version of the many-fermion problem is, in practice, not much different from the venerable random-phase approximation (*RPA*), albeit with some differences, mainly:

1.  The decoupling into independent sectors is not an approximation but is rigorous.

2.  Nor is anything "random", with the exception of ($\tilde{n}(\vec{k}) - <\tilde{n}(\vec{k})>$) which – when it appears under an integral sign – is subject to the Central Limit Theorem and therefore has to vanish.[7]

3.  Nonlinearities arise only indirectly, primarily from the dependence of the one-body distribution function on the coupling constant. For, otherwise, the Hamiltonian decouples neatly into a quadratic form in boson operators that we have identified as a set of individual "boson strings."

4. Asymptotic properties that can be calculated at small $|q|$ can be obtained with the least amount of effort, given that the effective *string lengths* are at a minimum if $|q|$ vanishes. This allows one to use results of the original Luttinger model, (string length =1) in studying the asymptotic sectors even in dimensions $d > 1$.

Obviously a lot of work that remains to be done. For example, one would wish to make firm connections with conventional many-body perturbation theory. One would also wish to express localized annihilation or creation *quasiparticle* operators $\Psi_\sigma(\vec{r})$ and $\Psi_\sigma^\dagger(\vec{r})$ as exponential functions of the various normal modes, as has been possible in 1D.[28]

But already at this stage the present author derives a great deal of satisfaction in demonstrating that a model of interacting fermions that existed initially only in 1D could illuminate the many-body problem in 2D and 3D. Or, to state it somewhat differently, on the sesquicentennial of Luttinger's discovery (or was it an invention?) we are happy – and somewhat astonished – to find it a platform still teeming with new possibilities.

---

**Appendix.** The formulation of this higher-dimensional extension of the canonical one-dimensional fermion gas interacting *via* short-range interactions owes its success to the absence or neglect of certain terms in the interaction Hamiltonian, an operator that is quartic in the field operators. (This is also true of the *RPA*, despite its many useful applications in the seventy years of *its* history.) We are, of course, referring to those terms that, in the one-dimensional models, are called "backward scattering" (they must also be treated separately) and to those that involve *umklapp*.[29] Leaving aside the latter because they are inconsequential in a low-density Fermi gas, let us look at the former. Among offending terms we find,

$$V(\vec{q})c^\dagger(\vec{k}-\vec{q})c^\dagger(\vec{k}'+\vec{q})c(\vec{k}')c(\vec{k}) \text{ where } \vec{k}\cdot\vec{q} > 0 \text{ and } |k| < |q|, \text{ while } \vec{k}'\cdot\vec{q} > 0 , \quad \text{(A.1)}$$

*i.e.*, operators that cannot be easily expressed in the boson picture.

The question is, do they invalidate the theory as it is worked out in the text? A quick answer is: not at small $q$ because this also implies small $k$, and for that there is an insignificant amount of phase space. Thus the theory, as it was derived in the text, remains valid *to some leading order in $q_0$*, the cutoff, regardless of these bad actors.

A deeper answer is: to the extent that such terms appear that change the occupation number by an *odd number* in any given sector, they do signal a failure of the picture of the gas of interacting fermions purely in terms of its normal modes. But there can be made an analogy to Grüneisen's parameter $\gamma$, a small number (order of 1/10) that measures the extent of nonlinearity in the elastic spectrum of ordinary solids. These are the nonlinearities that permit thermal expansion – impossible for an elastic solid – and give a finite lifetime to normal modes, thereby limiting their thermal conductivity to a finite value.

Logic requires that a picture of the normal modes be drawn first *before* investigating their bad behavior and that, of course, is what we have attempted to do here in the text.

affect the many-body eigenstates – when treated correctly. But with a cutoff at $q_0$ one gets around this objection while retaining the simplicity of a potential with $V(q) = U =$ constant – where it is not zero.

[13] The exclusion principle prevents two parallel spin fermions from residing on the same point. This is easy to see in coordinate space, using the identity $\tilde{n}_{j,\sigma}^2 = \tilde{n}_{j,\sigma}$ at each $\vec{R}_j$. The two-body interaction of Hubbard's model is $\frac{1}{2}U\sum_{\sigma,\sigma'}\tilde{n}_{j,\sigma}\tilde{n}_{j,\sigma'} = U\left(\tilde{n}_{j,\uparrow}\tilde{n}_{j,\downarrow} + \frac{1}{2}(\tilde{n}_{j,\uparrow}+\tilde{n}_{j,\downarrow})\right)$. After being summed over $j$ the two terms linear in the occupation operator sum up to a constant of the motion that is absorbed into the definition of the chemical potential (Fermi level) and, effectively, disappears. The remaining terms, quadratic in the occupation numbers, sum up to $H_2$ in the form shown in the text after Fourier transformation.

[14] In this sense they differ from the spin-wave operators in the Heisenberg model. Additionally, the attentive reader may notice that spin raising or lowering (but energy-lowering) operators $\sigma_0^\pm(\vec{q}) = x_0(\vec{q}) \pm iy_0(\vec{q})$ are easily constructed from Eqs.(17) and (18) or easily derived from the original fermion representation. Such operators have proved useful in various theories of magnetism, *cf.* D.C. Mattis, *Theory of Magnetism Made Simple,* World Scientific, Singapore, 2006, although in the present context trivial but annoying notational hurdles need to be overcome (*e.g.,* does one designate their energy-*raising* analogs $(\sigma_0^\pm)^\dagger$ ?) before they become either convenient or useful.

[15] J. Bardeen, L.N. Cooper and J.R. Schrieffer (BCS), Phys. Rev. **108**, 1175 (1957)

[16] Originally thought to be SO(4), see: C.N. Yang and S.C. Zhang, Mod. Phys. Lett. **B4**, 759-766 (1990)

[17] and not $-\delta_{\alpha,\beta}$ ! For these bosons to be properly normalized it is essential that $\varepsilon_1 + \varepsilon_2 > 0 \Rightarrow 1 - f(\varepsilon_1) - f(\varepsilon_2) > 0$ also; or that $\varepsilon_1 + \varepsilon_2 < 0 \Rightarrow 1 - f(\varepsilon_1) - f(\varepsilon_2) < 0$ also. These conditions are indeed met by the usual Fermi function $f(x)=(1+\exp(\beta x))^{-1}$ and even when (as we assume, with some degree of plausibility) two-body interactions affect the functional dependence of $f(\varepsilon)=<\tilde{n}(\varepsilon)>$; the only requirements are that $f(\varepsilon)$ be a monotonically decreasing function of $\varepsilon$ and that particle-hole symmetry imply $1-f(-\varepsilon)=f(\varepsilon)$.

[18] Unlike the $C$'s they depend only weakly on $q$, which is why we don't include $\sqrt{q}$ in their definition.